    \documentclass[12pt,draftclsnofoot,onecolumn]{IEEEtran}
\usepackage{epsfig,amsmath,amssymb,epsf,amsthm,scalefnt,multirow,subfig}
\usepackage{xcolor}
\usepackage{float}
\usepackage{cite}
\usepackage{psfrag}
\usepackage{bm}
\usepackage{algorithmic,algorithm}
\usepackage{graphicx}
\DeclareGraphicsExtensions{.pdf,.png,.jpg}

\newtheorem*{lemma*}{Lemma}

\begin{document}
\title{ Supervised-Learning for Multi-Hop MU-MIMO Communications with One-Bit Transceivers}
\author{\IEEEauthorblockN{Daeun Kim, Song-Nam Hong, and Namyoon Lee}\\
%\IEEEauthorblockA{
%Department of Electrical Engineering, POSTECH, \\Pohang, Gyeongbuk, Korea\\
%Emails: \{daeun.kim, somj163, nylee\}@postech.ac.kr and snhong@ajou.ac.kr} \\ 
\thanks{D. Kim and N. Lee are with the Department of Electrical Engineering, POSTECH, Pohang, Gyeongbuk 37673, South Korea  (e-mail: \{daeun.kim, nylee\}@postech.ac.kr).}
	\thanks{S.-N. Hong is with the Department of Electrical and Computer Engineering, Ajou University, Suwon, Gyeonggi 16499, South Korea (e-mail: snhong@ajou.ac.kr).}
}
  \maketitle
  \vspace{-1.2cm}
\begin{abstract}

This paper considers a nonlinear multi-hop multi-user multiple-input multiple-output (MU-MIMO) relay channel, in which multiple users send information symbols to a multi-antenna base station (BS) with one-bit analog-to-digital converters via intermediate relays, each with one-bit transceiver. To understand the fundamental limit of the detection performance, the optimal maximum-likelihood (ML) detector is proposed with the assumption of perfect and global channel state information (CSI) at the BS. This multi-user detector, however, is not practical due to the unrealistic CSI assumption and the overwhelming detection complexity. These limitations are addressed by presenting a novel detection framework inspired by supervised-learning. The key idea is to model the complicated multi-hop MU-MIMO channel as a simplified channel with much fewer and learnable parameters.  One major finding is that, even using the simplified channel model, a near ML detection performance is achievable with a reasonable amount of pilot overheads in a certain condition. In addition, an online supervised-learning detector is proposed, which adaptively tracks channel variations. The idea is to update the model parameters with a reliably detected data symbol by treating it as a new training (labelled) data. Lastly, a multi-user detector using a deep neural network is proposed. Unlike the model-based approaches, this model-free approach enables to remove the errors in the simplified channel model, while increasing the computational complexity for parameter learning. Via simulations, the detection performances of classical, model-based, and model-free detectors are thoroughly compared to demonstrate the effectiveness of the supervised-learning approaches in this channel.

%and shows  it is demonstrated that the proposed supervised-learning methods using a limited number of pilot symbols significantly outperforms the classical linear detection methods that even use  perfect and global CSI.
 \end{abstract}

%%%%%%%%%%%%%%% 
\section{Introduction}

Wireless relaying is an effective solution to expand network coverage and to enhance system reliability \cite{Boyer:04}. The use of multiple-input multiple-output (MIMO) systems is also a key technology for providing both considerable gains of spectral and energy efficiencies \cite{Larsson:14}. Motivated by these advantages, multi-user MIMO (MU-MIMO) relay networks have been considered as a promising cellular network architecture.  There have been extensive studies to characterize the capacity and to devise the effective communication schemes for the MU-MIMO relay networks over the past decade \cite{ Wang:05,Bolcskei:06,Tang:07,Fawaz:11,Jing:12,Rong:12,Yang:12}. In \cite{Wang:05,Bolcskei:06}, the information theoretical limits of the MIMO relay channels were characterized. It was shown in \cite{Yang:12} that the analytical expressions for outage probabilities were derived under a general channel fading distribution.  The underlying assumption of the aforementioned works, however, is that the relay and the BS equipped with multiple antennas use perfect hardware including infinite-precision analog-to-digital converters (ADCs) and digital-to-analog converters (DACs). When using a large number of antennas at the relay and the BS, the fabrication cost and the power consumption significantly increase. To diminish the power consumption and the cost, the use of cheaper and more energy-efficient hardware components including low precision ADCs and DACs has been considered as a promising approach \cite{Mezghani:07,Madhow:09,Wang:14,Studer:16, Hong:18,Jeon:TWC:18,Mo:18,Jeon:TVT:18}.  Motivated by this approach, this paper focuses on a multi-hop MU-MIMO relay channel, in which information symbols of users are delivered to the BS with one-bit ADCs via layered and distributed relays, each with one-bit transceiver. In this channel, it is very challenging to estimate the multi-hop channel accurately and detect information bits reliably because information symbols sent by the users are severely distorted by both multi-hop relays using one-bit transceivers and the BS using one-bit ADCs. In this paper, we present  novel supervised-learning approaches to reliably detect information symbols with a reasonable amount of pilots for channel training.

%In this setting, we first develop data detection algorithms from a classical communication system design point-of-view. To overcome the limitations of the classical communication system design approach, we take a novel machine learning approach by harnessing the intrinsic communication system model as side-information to accurately estimate channel state information and to reliably detect information symbols.
 
 \subsection{Related Works}
 
Despite the benefits of using low-precision ADCs and DACs at the relay and the BS, it changes not only the fundamental limits but also the required communication schemes including channel estimation and data detection. For single-hop communication networks in which the multi-antenna BS employs the low-precision ADCs, the channel estimation and data detection algorithms have been proposed in \cite{Wang:14,Studer:16, Hong:18,Jeon:TWC:18,Mo:18,Jeon:TVT:18}. Recently, asymptotic achievable rates have been characterized for dual-hop MU-MIMO systems when the low-precision ADCs are used at either the relay \cite{Dong:17,Kong:18} or the  BS \cite{Liu:17}. The key tool for the analysis of the achievable rates in \cite{Dong:17,Kong:18,Liu:17} is the use of the additive quantization noise model (AQNM) by leveraging the Bussgang's decomposition \cite{Bussgang}. In \cite{Cao:17}, the channel estimation methods using support vector machine and neural networks were proposed when the one-bit relay cluster was considered. The common limitation of the prior works is that they consider the one-bit quantization at either the relay or the BS. Therefore, the joint impact of low-resolution ADCs in the two-hop relying system is still unknown. In addition, the existing works focused on the single-hop relay networks; thereby, the effects of channel estimation and data detection when scaling the number of hops are also unrevealed.

 There have been increasing research interests in exploiting machine learning tools to address the nonlinearity of a MIMO system with low-resolution ADCs. By treating an end-to-end nonlinear MIMO system with low-resolution ADCs as an autoencoder, a supervised-learning aided communication framework was proposed in \cite{Jeon:TVT:18}. Specifically, it empirically learns the nonlinear channel (i.e., the conditional probability mass functions (PMFs)) by sending pilot symbols (or known data symbols) repeatedly. Leveraging the learned channel, novel empirical ML-like and minimum-center-distance detectors were proposed. Following this work, a reinforcement learning aided detector was presented in \cite{Jeon:WCNC:18}, in which a cyclic redundancy check (CRC) code is used to obtain a new labelled data set to further improve the estimation accuracy of the PMFs. Recently, in \cite{Andrews:18},  a deep-learning detector was also proposed for an orthogonal frequency division multiplexing (OFDM) system using one-bit ADCs, which can address the nonlinear distortion caused by one-bit quantizations.  Beyond the nonlinearity induced by one-bit ADCs, in \cite{Farsad,Hoydis,Hoydis2,Hoydis3}, numerous deep-learning based joint detection and decoding methods were proposed for linear/nonlinear channels by treating an end-to-end communication system as an autoencoder. To our best knowledge, however, all the aforementioned machine learning based channel-training and data detectors have not been considered for a nonlinear multi-hop MU-MIMO relay channel, which involves multiple nonlinear one-bit quantization effects in a cascade manner.

%%%%%%%%%%%%%
 \subsection{Contributions}
 
In this paper we focus on a nonlinear multi-hop MU-MIMO relay channel, in which $K$ single-antenna users transmit data symbols to the BS equipped with $N$ antennas with the help of the $M$ layered and distributed relays. The nonlinearity of this channel comes from the assumption that both the relays and the BS use one-bit ADCs, and the relays also use one-bit DACs for transmissions. The major contributions of this paper are summarized as follows.

 %comprised of four parts: (i) the classical data detection methods, (ii) model-based end-to-end supervised-learning for channel training and data detection, (iii) model-based end-to-end online supervised-learning for channel training and data detection, and (iv) model-less end-to-end supervised-learning via deep neural networks for channel training and data detection.
  
 {\bf Classical communication approach:} Inspired by the classical approach of a communication system, we first derive the optimal maximum-likelihood (ML) detector for the nonlinear multi-hop MU-MIMO relay channel, by assuming that the BS has global and perfect knowledge of channel state information (CSI). This ML detector provides the fundamental limit of the detection performance in the channel. Toward this, we characterize the end-to-end transition probability distribution of the multi-hop channel as a function of a per-hop signal-to-noise ratio (SNR) and CSI. In practice, however, the use of the derived ML detector is impossible, because acquiring global and perfect CSI at the BS is infeasible even using an infinite number of pilots. Moreover, the computational complexity of the ML detector increases exponentially with both the number of hops and relays per hop. Because of such limitations, it is pessimistic to apply the classical communication approach for the nonlinear multi-hop MIMO channels.

 %This is because the number of pilots for the channel parameter estimation is limited per hop. In addition, even using an infinite number of pilots, the accurate CSI estimation is impossible due to both high quantization noise effects and the distributed nature of relays using one-bit transceivers. 

 %Although the derived ML detector is optimal in the sense of minimizing detection errors, it cannot be applicable in practice, because it requires global and perfect CSI. Acquiring this CSI knowledge, however, is impossible even using an infinite number of pilots due to severe quantization errors by one-bit ADCs. Another limitation of the ML detector is its prohibitive computational complexity, which exponentially increases with both the number of hops and the number of relays per hop, i.e., $\mathcal{O}\left(|\mathcal{ \tilde C}|^{KML}\right)$, where $\mathcal{ \tilde C}$ is a constellation set of information symbols. 
 
 %To reduce the computational complexity, we also propose a multi-stage linear minimum mean square error (LMMSE) detection method, which successively applies the Bussgang decomposition hop-by-hop. It is shown that the computational complexity of this linear detection metper hod scales with both the number of hops linearly and the number of relays per hop quadratically, i.e., $\mathcal{O}\left(ML^2\right)$. 
  
  {\bf Model-based supervised-learning approach:} To overcome the limitations of the classical approach, we propose a novel communication framework using a model-based supervised-learning approach. Unlike the model-free approach via deep learning in \cite{Farsad,Hoydis,Hoydis2,Hoydis3}, the proposed framework is to model the end-to-end multi-hop MU-MIMO channel as simple $2N$ parallel binary symmetry channel (BSCs), which can be characterized by much fewer and learnable model parameters than those in the original channel. The parameters of the effective BSCs include 1) a set of $2N$-dimensional binary vectors (i.e., codewords) and 2) the crossover probabilities of the BSCs. During a training phase, the parameters of the effective channel model are jointly trained with a reasonable number of pilots. Subsequently, during the phase of data transmission, the BS performs a weighted minimum Hamming distance detection (wMHD) to recover the information symbols using the estimated model parameters. We call this as an approximate ML (A-ML) detector.  To verify the effectiveness of the proposed framework, we show that the A-ML detector achieves a near ML performance even with a reasonable amount of pilots, provided that the SNRs of the previous $M-1$ hops are sufficiently high. One major observation is that this model-based approach reduces the number of parameters to learn in the complex nonlinear MIMO channel compared to the classical approach. As a result, the detection performance of the model-based approach significantly outperforms the classical approach for a given pilot overhead.

{\bf Model-based online-learning approach:} Despite its attractive performance, the proposed supervised-learning framework is the lack of flexibility in adapting environment changes. For instance, when a channel value of a certain hop is time-varying, the model parameters including the codebook and the crossover probabilities should be updated accordingly, since they are the function of the multi-hop channels. We further improve the proposed supervised-learning framework so that it is robust to time-varying channel environments. The proposed framework jointly performs the model parameter update and the data detection, similar to an expectation-maximization (EM) algorithm. Specifically, during a data transmission phase, the BS first assigns a label to the received signal vector by computing a posteriori probability (APP) of it. Then, this APP information is exploited to estimate and update the model parameters. Subsequently, using the updated model parameters, the BS performs data detection. Our key finding is that the proposed online-learning approach can outperform the conventional linear detectors using genie-aided (perfect and global) CSI at the BS under some time-varying channel conditions.

%To adaptively tract channel variations, we also present a model-based end-to-end online supervised-learning method for channel training-tracking and data detection. Inspired by an expectation-maximization (EM) algorithm, the proposed online-learning algorithm iteratively performs the update of the model parameters and data detection. Specifically, during the data transmission phase, the BS first computes the reliability of each information symbol vector by using the previously estimated model parameters. Then, using the reliability information, the algorithm performs the update of the model parameters and detection. Our key finding is that the proposed online-learning approach is very robust to the time-varying multi-hop MU-MIMO systems. 
  	
 {\bf Model-free deep learning approach:}  The proposed model-based supervised-learning approaches are very effective when training and tracking the channel model parameters. This is because the number of the model parameters for channel training is much fewer than that in the original channel model. In addition, the parameters in the proposed model are accurately estimated by a simple training strategy compared to those in the original model.  These approaches, however, are fundamentally limited when the model contains a modeling error. Since the effective BSC channel model is a good approximation of the original channel model when the $M-1$ hop SNRs are high enough, the model-based learning approaches can degrade the performance when the $M-1$ hop SNRs are low. To resolve this model error problem, we also propose a multi-user detector via a deep neural network (DNN) using a model-free approach. To be specific, we construct a DNN comprised of the multiple layers, and optimize the parameters of the DNN by sending a few repetitions of data symbols as pilots. One noticeable observation is that this approach can improve the detection performance by eliminating the model error in some cases. Nevertheless, this approach is not suitable for the scenarios where the channel changes relatively fast, because the computational complexity for training the DNN parameters is very high compared to the model-based approaches.

   \begin{figure*}[t]
 \center
  \includegraphics[width=12cm]{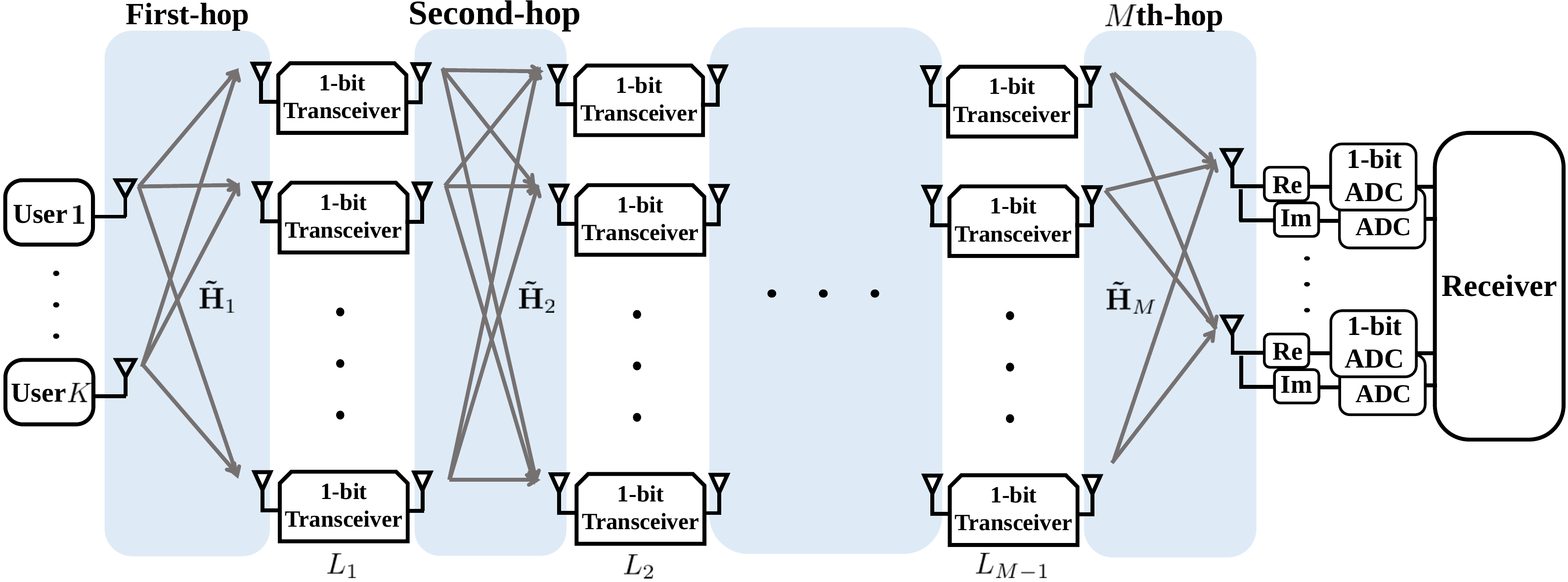}\vspace{-0.3cm}
  \caption{An illustration of the nonlinear $M$-hop MU-MIMO relay channel.} \label{Fig1}
\end{figure*}  

%%%%%%%%%%%%%
\section{System model}
In this section, we consider a nonlinear $M$-hop MU-MIMO relay channel.  As illustrated in Fig. \ref{Fig1}, $K$ users send information symbols to a BS with the aid of $M-1$ layered and distributed relays, each with one-bit transceiver. We assume that the BS is equipped with $N$ antennas, each with one-bit ADCs.

{\bf First-hop transmission:} Let ${\tilde x}_{k}$ be an information symbol of the $k$th user, which is chosen from a constellation set $\mathcal{M}_{\rm t}$. In addition, let $\mathbf{\tilde x}=[{\tilde x}_{1},{\tilde x}_{2},\ldots, {\tilde x}_K]^{\top} \in \mathcal{M}_{\rm t}^{K \times 1}$ denote the aggregated data symbol vector sent by all $K$ users. We denote the complex channel from the $k$th uplink user to the $\ell$th relay at the first-hop by ${\tilde h}_{1,\ell,k}$. Then, the received signal of the $\ell$th relay with one-bit ADCs at the first hop is 
\begin{align}
	{\tilde r}_{1,\ell} = {\sf sign}\left(\sum_{k=1}^K{\tilde h}_{1,\ell,k}{\tilde x}_{k}+{\tilde v}_{1,\ell}\right),
\end{align}
where ${\tilde v}_{1,\ell}$ is a circularly-symmetric complex Gaussian random variable with zero mean and variance $\sigma^2_{1}$, i.e., ${\tilde v}_{1,\ell}\sim \mathcal{CN}(0,\sigma^2_{1})$ and ${\sf sign}(\cdot):\mathbb{R}\rightarrow \{-1,+1\}$ denotes the one-bit quantization function, which is independently applied to the real and imaginary components.% of $\sum_{k=1}^K{\tilde h}_{1,\ell,k}x_{k}+{\tilde v}_{1,\ell}$. 

{\bf Relay operation:} Since each relay is assumed to equip with one-bit DACs, it transmits a quadrature phase shift keying (QPSK) symbol for the second-hop transmission. Let ${\tilde s}_{m,\ell}\in \mathcal{M}_r$ be the transmit signal of the $\ell$th relay at the $m$th hop where $\mathcal{M}_r=\frac{1}{\sqrt{2}}\left\{1+j,1-j,-1+j,-1-j\right\}$ with $|\mathcal{M}_r|=4$. In particular, the relay transmission symbol is constructed by
\begin{align}
{\tilde s}_{m,\ell}=	f_{\rm r}\left({\tilde r}_{m,\ell} \right)\in \mathcal{M}_r,
\end{align}
 where $f_{\rm r}(\cdot): \mathcal{M}_r\rightarrow \mathcal{M}_r$ denotes a relay operation function that uniquely maps a received signal of the relay to its transmit signal. For simplicity, we assume a relay operation function which simply forwards the binary received signal to the next hop, i.e., $f_{\rm r}\left({\tilde r}_{m,\ell} \right)={\tilde r}_{m,\ell}$.
%  For example, a simple relay operation function is to send the same signal with the received signal, i.e., $f_{\rm r}\left({\tilde r}_{\ell} \right)={\tilde r}_{\ell}$, which is a reasonable relaying strategy when the first-hop SNR is sufficiently large.

{\bf Multi-hop transmission:}
We denote the number of relays in the $m$th layer by $L_m$ for $m\in \{1,2,\ldots, M-1\}$. We also denote the channel from the $\ell$th relay transmission of the $m$th hop by ${\bf {\tilde h}}_{m,\ell}\in \mathbb{C}^{N\times 1}$. Then, the received signal of the relays with one-bit ADCs at the $m$th hop is
\begin{align}
	{\bf \tilde r}_{m}&={\sf sign}\left(\sum_{\ell=1}^{L_{m-1}}{\bf \tilde h}_{m,\ell}{ \tilde r}_{m-1,\ell}+{\bf \tilde v}_{m}\right), 
\end{align} 
where ${\mathbf{ \tilde v}_{m}}=[{\tilde v}_{m,1},\ldots, {\tilde v}_{m,L_m}]^{\top}\in \mathcal{C}^{L_m\times 1}$ is the noise vector at the $m$th hop. The elements of ${\mathbf{\tilde v}_m}$ are independent and identically distributed (IID) complex Gaussian random variables, i.e., ${\tilde v}_{m,\ell}\sim \mathcal{NC}(0,\sigma^2_{m})$.
Considering the $M$-hop relaying systems, the received signal at the BS is given by
\begin{align}
    {\bf \tilde y}={\sf sign}\left(\sum_{\ell=1}^{L_{M-1}}{\bf \tilde h}_{M,\ell}{ \tilde r}_{M-1,\ell}+{\bf \tilde v}_{M}\right). \label{eq:received_BS}
\end{align}
Let ${\mathbf{ \tilde H}_m}=[{\mathbf{\tilde h}}_{m,1},\ldots, {\bf \tilde h}_{m,L_{m-1}}]\in \mathbb{C}^{L_m\times L_{m-1}}$ be the channel matrices of the $m$th hop. Then, the received signal of the BS in \eqref{eq:received_BS} can be written in a matrix form  as
\begin{align}
	{\bf \tilde y}&={\sf sign}\left({\bf \tilde H}_M {\bf \tilde r}_{M-1} +{\bf \tilde v}_M\right). \label{eq:received_BS_vec}
\end{align}

% \begin{figure}[t]
%     \centering
% \epsfig{file=twohop_1.pdf,width=7.0cm}
% \caption{Two-hop multi-user MIMO systems with $L$ relays.} \label{Fig1}
% \end{figure}

For the notational simplicity, we rewrite the input and output relationship in \eqref{eq:received_BS_vec} into a real-representation as
\begin{align}
	{\bf y}&={\sf sign}\left({\bf H}_M \mathbf{r}_{M-1}+{\bf  v}_M\right),\label{eq:received_BS_vec_real}
\end{align}
where ${\bf y}=\left[{\sf Re}({\bf \tilde y})^{\top}, {\sf Im}({\bf \tilde y})^{\top}\right]^{\top} $, ${\bf r}_{M-1}=\left[{\sf Re}({\bf \tilde r}_{M-1})^{\top}, {\sf Im}({\bf \tilde r}_{M-1})^{\top}\right]^{\top}$, ${\bf v}_M=\left[{\sf Re}({\bf \tilde v}_M)^{\top}, {\sf Im}({\bf \tilde v}_M)^{\top}\right]^{\top}$, and
$    {\bf  H}_M = 	\begin{bmatrix} 
   {\sf Re}({\bf \tilde H}_M)& -{\sf Im}({\bf \tilde H}_M)\\ {\sf Im}({\bf \tilde H}_M)&{\sf Re}({\bf \tilde H}_M)\end{bmatrix}. $
This real-representation can be applied to each hop straightforwardly and will be used in the sequel. 
\section{ML Detection using Classical Approach}
In this section, from a classical communication system design point-of-view, we propose a ML detector when the global CSI is perfectly known to the BS.  Although this CSI assumption is unrealistic, the ML detector can provide the fundamental limit of the detection performance in this network. %In addition, we provide an analytical expression of the symbol-error-vector probability (SEVP) of the ML detection. 

%The first one is the optimal ML detection algorithm. Then, to reduce the computational complexity, we also propose a LMMSE detector by applying the Bussgang decompositions successively.

%\subsection{ML Detection}

To derive the ML detector, we need to characterize the effective channel transition probabilities for a given channel input vector. To accomplish this, we define $M+1$ channel input and output sets in the relay network. Let $\mathcal{X}=\{{\bf x}_1,{\bf x}_2,\ldots, {\bf x}_{|\mathcal{M}_{\rm t}|^{K}}\}$ denote the channel input set containing all  possible transmitted vectors by the $K$ users, i.e., ${\bf x} \in \mathcal{X}$ and $|\mathcal{X}|=|\mathcal{M}_{\rm t}|^{K}$ where $|\mathcal{M}_{\rm t}|$ is the constellation size. We also define the channel output and input sets of the $L_m$ distributed relays with one-bit ADCs by $\mathcal{R}_m=\{{\bf r}_{m,1},{\bf r}_{m,2},\ldots, {\bf r}_{m,2^{2L_m}}\}$, where $m\in\{1,2,\ldots,M-1\}$. Similarly, the channel output set of the BS is defined by $\mathcal{Y}=\{{\bf y}_1,{\bf y}_2,\ldots, {\bf y}_{2^{2N}}\}$. Using these sets, we  compute the channel transition probabilities of the $M$-hop. We first consider the first-hop channel transition probability. The probability that the received signal vector of the $L_1$ relays is ${\bf r}_{1,u}$ when the $K$ uplink users transmit ${\bf x}_i$ is computed as
 \begin{align}
 	 \mathbb{P}\left[ {\bf r}_1={\bf r}_{1,u}| {\bf x}={\bf x}_{i}\right] 
 	& =   \prod_{\ell=1}^{2L_1}\mathbb{P}\left[ {r}_{1,\ell}={r}_{1,u,\ell}| {\bf x}={\bf x}_{i}\right] \nonumber\\ 
 	& =   \prod_{\ell \in \mathcal{Z}_u^{+}}\mathbb{P}\left[   {\bf h}_{1,\ell}^{\top}{\bf x}_{i} + v_{1,\ell} > 0\right]  \prod_{\ell \in \mathcal{Z}_u^{-}}\mathbb{P}\left[ {\bf h}_{1,\ell}^{\top}{\bf x}_{i} + v_{1,\ell} < 0\right]  \nonumber\\ 
 		& =  \prod_{\ell=1}^{2L_1} Q\left(\frac{- r_{1,u,\ell}  {\bf h}_{1,\ell}^{\top}{\bf x}_{i}}{\mathbf{\sigma}_{1}/ \sqrt{2}}\right ),\label{eq:firsthop_CTP}
 \end{align}
 where $\mathcal{Z}_u^+=\{\ell| r_{1,u,\ell}=1\}$ and $\mathcal{Z}_u^-=\{\ell| r_{1,u,\ell}=-1\}$ are the sets indicating the sign of the $\ell$th element of ${\bf r}_{1,u}$. Here, $Q(x)=\int_{x}^{\infty} \frac{1}{\sqrt{2\pi}}e^{-\frac{u^2}{2}}{\rm d} u$ is the standard Q-function. For the $m$th-hop channel for $m\in \left\{2,\ldots,M-1\right\}$, the received signal vector of the relays at the $m$th hop is $\mathbf{r}_{m,w}$ when the previous relays transmit $\mathbf{r}_{m-1,v}$ is computed as
 \begin{align}
     \mathbb{P}\left[ {\bf r}_m={\bf r}_{m,w}| {\bf r}_{m-1}={\bf r}_{m-1,v}\right]   &= \!  \prod_{j=1}^{2L_m}\mathbb{P}\left[ {r}_{m,j}={r}_{m,w,j}| {\bf r}_{m-1}={\bf r}_{m-1,v}\right] \nonumber\\ 
 	& =\!\!   \prod_{j \in \mathcal{Z}_w^{+}}\!\!\mathbb{P}\!\left[   {\bf h}_{m,j}^{\top}{\bf r}_{m-1} \!+\! v_{m,j} \!>\! 0\right]  \!\!\!\prod_{j \in \mathcal{Z}_w^{-}}\!\!\mathbb{P}\!\left[ {\bf h}_{m,j}^{\top}{\bf r}_{m\!-\!1,v} \!+\! v_{m,j} \!<\! 0\right]  \nonumber\\ 
 		& =  \prod_{j=1}^{2L_m} Q\left(\frac{- r_{m,w,j}  {\bf h}_{m,j}^{\top}{\bf r}_{m-1,v}}{\mathbf{\sigma}_{m}/ \sqrt{2}}\right ).
 \end{align}
 For the $M$th-hop channel, the probability that the received signal vector of the BS is ${\bf y}_j$ when the relays transmit ${\bf r}_{M-1,u}$ is computed as
 \begin{align}
 	 \mathbb{P}\left[ {\bf y}={\bf y}_j| \mathbf{r}_{M-1}={\bf r}_{M-1,u}\right]  
 	& =   \prod_{n=1}^{2N}\mathbb{P}\left[ {y}_{n}={y}_{j,n}| \mathbf{r}_{M-1}={\bf r}_{M-1,u}\right] \nonumber\\ 
 	& =   \!\prod_{n\in \mathcal{S}_j^{+}}\!\mathbb{P}\!\left[   {\bf h}_{M,n}^{\top}{\bf r}_{M-1,u} + v_{M,n} > 0\right]  \!\prod_{n\in \mathcal{S}_j^{-}}\!\!\mathbb{P}\!\left[   {\bf h}_{M,n}^{\top}{\bf r}_{M-1,u}+ v_{M,n} < 0\right]  \nonumber\\ 
 		& =  \prod_{n=1}^{2N} Q\left(\frac{- y_{j,n}  {\bf h}_{M,n}^{\top}{\bf r}_{M-1,u}}{\mathbf{\sigma}_{M}/ \sqrt{2}}\right ), \label{eq:finalhop_CTP}
 \end{align}
 where $\mathcal{S}_j^+=\{n| y_{j,n}=1\}$ and $\mathcal{S}_j^-=\{n| y_{j,n}=-1\}$ are the index sets. From \eqref{eq:firsthop_CTP} to \eqref{eq:finalhop_CTP}, the probability that the received signal of the BS is ${\bf  y}_j$ when the uplink transmission vector is given by ${\bf x}_i$ is computed as a sum-product form of the standard Q-function, namely, 
\begin{align}
 \mathbb{P}\left[{\bf y}={\bf y}_j|{\bf x}={\bf x}_i\right] 
&=  \sum_{ {\bf r}_{M-1,u} \in \mathcal{R}_{M-1} }\ldots \sum_{ {\bf r}_{1,u} \in \mathcal{R}_{1} }
\mathbb{P}\left[ {\bf y}={\bf y}_j| {\bf r}_{M-1}={\bf r}_{M-1,u}\right]  \times\cdots \times \mathbb{P}\left[ {\bf r}_{1}={\bf r}_{1,u} | {\bf x}=\mathbf{x}_i \right] \nonumber\\
&=\!\!  \sum_{ {\bf r}_{M-1,u} \in \mathcal{R}_{M-1} }\!\!\ldots\!\! \sum_{ {\bf r}_{1,u} \in \mathcal{R}_{1} }
 \prod_{n=1}^{2N}\! Q\!\left(\!\frac{- y_{j,n}  {\bf h}_{M,n}^{\top}{\bf r}_{M-1,u}}{\mathbf{\sigma}_{M}/ \sqrt{2}}\!\right )\times \!\cdots \!\times \! \prod_{\ell=1}^{2L_1} \!Q\!\left(\frac{- r_{1,u,\ell}  {\bf h}_{1,\ell}^{\top}{\bf x}_{i}}{\mathbf{\sigma}_{1}/ \sqrt{2}}\right )\!. \label{eq:likelihood}
   \end{align} 
  From the effective channel transition probabilities between ${\bf x}_i\in \mathcal{X}$ and ${\bf y}_j\in \mathcal{Y}$, the optimal ML detector is  
  \begin{align}
    g^{{\sf ML}}({\bf y}_j)=\arg\max_{{\bf x}_i\in \mathcal{X}} \mathbb{P}\left[{\bf y}={\bf y}_j|{\bf x}={\bf x}_i\right].  \label{eq:ML_detection}
  \end{align}

%   \begin{align}
% &g^{{\sf ML}}({\bf y}_j) \nonumber \\
% & \!= \!\arg\max_{{\bf x}_i\in \mathcal{X}}\! \sum_{ {\bf r}_{u}\! \in \mathcal{R} }
% \! \prod_{n=1}^{2N} \!Q\!\left(\!\!\frac{- y_{j,n}  {\bf g}_n^{\top}f_{\rm r}({\bf r}_u)}{\mathbf{\sigma}_w/ \sqrt{2}}\!\right )\!\!\prod_{\ell=1}^{2L}\! Q\!\left(\!\!\frac{- r_{u,\ell}  {\bf h}_{\ell}^{\top}{\bf x}_{i}}{\mathbf{\sigma}_v/ \sqrt{2}}\!\!\right )\!. \label{eq:ML_detection}
% \end{align}

We explain some remarks on the ML detector in \eqref{eq:ML_detection}.

 {\bf Remark 1 (Need for global and perfect CSI):} As derived in \eqref{eq:ML_detection}, the BS needs to know 1) global and perfect CSI of the network at the BS, i.e., $\left\{{\bf H}_m\right\}_{m=1}^M$, and 2) all possible realizations of the received signal at the relays, ${\bf r}_m \in \mathcal{R}_{m}$ to perform the optimal ML detection. Specifically, the number of parameters (unknowns) for the channel estimation quadratically increases with the number of hops, i.e., $\mathcal{O}(2KNML^2)$, where $L_m=L$ for $\forall m$.  Since the relays and the BS equip with one-bit ADCs, it is infeasible to obtain the accurate CSI from conventional pilot transmission methods. %Consequently, devising a detection algorithm that is robust to CSI errors is important in practice.
 
 %Consequently, a ML detector robust to CSI errors is required. In addition, the BS requires to know the relaying functions $f_{\rm r}({\bf r}_u)$ for all ${\bf r}_u \in \mathcal{R}$ and the number of possible relaying functions exponentially increases with the number of the relays. Therefore, the design of the simple relaying function is an important issue.
 
   %$f_{\rm r}\left({\tilde r}_{\ell} \right)={\tilde r}_{\ell}$ is considered in the sequel. 

%{\bf Remark 1:} As derived in \eqref{eq:likelihood}, the BS needs to know the global and perfect CSI of the network, i.e., $\mathbf{H}_1 \ldots \mathbf{H}_M$. Since the relays and the BS equip with one-bit ADCs, it is difficult to obtain the accurate CSI from conventional pilot transmission methods. Consequently, a ML detector that is robust to CSI errors is required.

{\bf Remark 2 (Computational complexity):} For the single-hop multi-user MIMO system with one-bit ADCs, it has shown in \cite{Wang:14} that the ML detection problem can be solved in a computationally efficient manner by convex relaxation techniques with the logarithmically-concave property of the likelihood function. Whereas, the convex optimization algorithms cannot be applied in the multi-hop relaying system, because the likelihood function in \eqref{eq:likelihood} is neither concave nor logarithmically-concave. The detection computational complexity exponentially increases with the number of uplink users, the relays per layer, and hops, i.e., $\mathcal{O}\left(|\mathcal{M}_{\rm t}|^{K}2^{\sum_{m=1}^{M-1}2L_m}\right)$. This computational complexity hinders the use of the ML detector in practice. 

\section{Model-based Supervised-Learning Approach} \label{modelbased}
In this section, we propose a novel communication framework for the nonlinear multi-hop MU-MIMO relay channel by harnessing an end-to-end supervised-learning technique.  We first present a simple model that can be a good approximation of the complicated nonlinear multi-hop MU-MIMO relay channel by exploiting a coding theoretical framework developed in our prior works \cite{Hong:18, Jeon:TVT:18}.  Then, we explain how to learn the model parameters using a simple training strategy and to detect the data symbols using the trained model parameters. In addition, we prove that the proposed channel training and data detection framework can achieve the optimal ML detection performance under certain conditions.  

%, the proposed strategy is to jointly learn the parameters, and use them for the data detection. 
%The , 

%not only accurately captures the nonlinear network, but also is easily  trainable with a reasonable amount of parameters. 
%
% ML (A-ML) detection method based on pilot transmissions, i.e., imperfect CSI at the BS.  The key idea of the proposed method is to learn both the codebook and the empirical channel transition probabilities for the A-ML detection by using both a coding-theoretical approach proposed and the end-to-end learning. 

\subsection{The Proposed End-to-End Network Model}

The data detection problem for the nonlinear multi-hop MU-MIMO relay channel can be reformulated as a decoding problem of channel-dependent nonlinear codes from a coding-theoretical perspective \cite{Hong:18}. To explain this method, we first introduce the notions of the codebook construction and the BSC model for the corresponding decoding problem.
% which is equivlaent to the data detection problem for the dual-hop multi-user MIMO relaying systems with one-bit ADCs.  

{\bf Codebook construction:} Let us define a codeword vector ${\bf c}_i$, which is generated by an encoding function $f({\bf x}_i,{\bf H}_m)$ that maps information vector ${\bf x}_i\in \{-1,+1\}^{2K}$ into an $2N$-dimensional binary space $\{-1,+1\}^{2N}$. In particular, the encoding function is given by 
\begin{align}
	{\bf c}_i&={\sf sign}\left({\bf H}_M {\sf sign}\left({\bf H}_{M-1}\cdots{\sf sign}\left(\mathbf{H}_1{\bf  x}_i\right) \right)\right).\label{eq:encoding}
\end{align}
As can be seen in \eqref{eq:received_BS_vec_real}, this codeword vector is a noise-free binary representation of the received signal when the uplink users transmit ${\bf x}_i$. We also define a codebook as the collection of all possible codeword vectors as $\mathcal{C}=\left\{{\bf c}_1,{\bf c}_2,\ldots, {\bf c}_{|\mathcal{M}_{\rm t}|^{K}}\right\}$. The cardinality of $\mathcal{C}$ is less than or equal to that of $|\mathcal{X}|$, i.e., $|\mathcal{C}|\leq |\mathcal{X}|$. This is because for a certain realization of ${\bf H}_m$, it is possible that ${\bf c}_i={\bf c}_j$ for two distinct information vectors ${\bf x}_i$ and ${\bf x}_j$ where $i\neq j$. 
 In addition, since $2K$ binary information bits are sent over $2N$ channel uses, a code rate of this encoding function is $r=\frac{2K}{2N}=\frac{K}{N}$, which is typically less than 1/2 for a massive MIMO setting.  Furthermore, the codes are not a linear class, i.e., a linear combination of two codewords is not necessarily in $\mathcal{C}$.

% \begin{figure*}
%  \center
%   \includegraphics[width=\textwidth]{training.pdf}
%   \caption{ML decoding via the minimum weighted Hamming distance decoding (MWHD).}
% \end{figure*} 

{\bf Channel parameters:} When a codeword ${\bf c}_i\in \{-1,+1\}^{2N}$ is generated by the encoding function in \eqref{eq:encoding}, it is sent over noisy channels. Then, the BS receives ${\bf y}\in \{-1,+1\}^{2N}$. Since noise signals over different BS antennas are independent, the channel between ${\bf c}_i$ and ${\bf y}$ can be  modeled by $2N$-parallel BSCs, each with different crossover probabilities, i.e.,  
\begin{align}
	\mathbb{P}({\bf y}|{\bf c}_i) =\prod_{n=1}^{2N}\mathbb{P}(y_n|c_{i,n}),
\end{align}
where \begin{align}
	\mathbb{P}(y_n|c_{i,n})=  \begin{cases}
p_{i,n} & {\rm if}~  y_n \neq c_{i,n},\\
1-p_{i,n} &{\rm if}~ y_n = c_{i,n}.
\end{cases}
\end{align}
Notice that the effective channel crossover probabilities $\{p_{i,n}\}_{n=1}^{2N}$ varies over both the channel uses and the channel input. This is a major difference with the classical coding problem setting.

{\bf  Remark 3 (Our modeling and limitation):} We converted the nonlinear multi-hop MU-MIMO relay channel into a simple channel model comprised of $2N$ BSCs. This simplified model is parameterized by the codebook $\mathcal{C}$ and the set of transition probabilities $\{p_{i,n}\}_{n=1}^{2N}$. Therefore, during the training phase, these parameters should be trained to fit our model with the labeled training data set, i.e., a sequence of pilot symbols. Although our parallel BSC model is simple, it cannot capture the propagation effects of the correlated noise signals in the multi-hop channels. Nevertheless, it turns out that our simple model can be optimal in the sense of minimizing detection error for the case when the first $M-1$ hop SNRs are sufficiently high.  This optimality result will be provided in Section IV-C.

%These cases can be   fact can be facilitated in the practical multi-hop MU-MIMO relay network using one-bit transceivers.
  
%{\bf Weighted Hamming Distance:} For the given codebook $\mathcal{C}$, the distance between a codeword vector ${\bf c}_i=[c_{i,1}, \ldots, c_{i,2N}]^{\top}$ and the received signal vector ${\bf y}=[y_{1}, \ldots, y_{2N}]^{\top}$ are measured by the weighted Hamming distance, i.e.,
%\begin{align}
%	d_{\sf wH}({\bf c}_i,{\bf y};{\bf \alpha}, {\bf \beta}) \!=\!\sum_{n=1}^{2N}\! \alpha_n|c_{i,n}\!-\!y_{n}|_0 +\beta_n(1\!-\!|c_{i,n}\!-\!y_{n}|_0),
%\end{align}
%where ${\bf \alpha}=[\alpha_1,\ldots,\alpha_{2N}]$ and ${\bf \beta}=[\beta_1,\ldots,\beta_{2N}]$ are the weight vectors and the elements of them are defined by the crossover probabilities.

%\subsection{End-to-End Training and Data Detection}
%In this subsection, we explain the training for model parameter estimation and the data detection method via MWHD using the estimated model parameters.  

%{\bf End-to-end training by pilot transmisT_{\rm d}sion:} During the training phase, $K$ users sends  

\subsection{Parameter Learning and Detection Algorithm}
 Each transmission frame containing $T_{\rm B}$ times slots consists of two phases: 1) a channel training phase with $T_{\rm t}$ time slots and 2) a data transmission phase with $T_{\rm d}$ time slots, i.e., $T_{\rm B}=T_{\rm t}+T_{\rm d}$. 
 
  {\bf Acquisition of training examples:} Let $T$ be the number of repetitions for training of each ${\bf x}_i \in \mathcal{X}$ where $|\mathcal{X}|=|\mathcal{M}_{\rm t}|^{K}$. During the channel training phase, $K$ uplink users repeatedly send each information vector ${\bf x}_i \in \mathcal{X}$, i.e., ${\bf x}[t]={\bf x}_i$ for $t\in \mathcal{T}_i =\{ t: (i-1)T+1\leq t\leq iT\}$.  As a result, a total of $T|\mathcal{M}_{\rm t}|^{K}$ time slots is required during the training phase, i.e., $T_{\rm t}=T|\mathcal{M}_{\rm t}|^{K}$. Let ${\bf X}_t=\left[{\bf x}[1],{\bf x}[2], \ldots, {\bf x}[T_{\rm t}]\right] \in \{-1,+1\}^{2K\times T_{\rm t}}$ and ${\bf Y}_t=\left[{\bf y}[1],{\bf y}[2], \ldots, {\bf y}[T_{\rm t}]\right] \in \{-1,+1\}^{2N\times T_{\rm t}}$ be the sets of transmit and received signal vectors during the training phase. We define $\mathcal{S}=\left\{ ({\bf x}[1],{\bf y}[1]), \ldots, ({\bf x}[T_{\rm t}],{\bf y}[T_{\rm t}])\right\}$ as a set of labeled training examples. 
  
 {\bf Codebook learning:} Using $\mathcal{S}$, the BS first estimates codebook $\mathcal{\hat C}=\{{\bf \hat c}_1,\ldots, {\bf \hat c}_{|\mathcal{M}_{\rm t}|^{K}}\}$. To accomplish this, during the training phase, the BS computes the centroid vector when transmitting ${\bf x}_i$ using the received samples $\{{\bf y}[(i\!-\!1)T\!+\!1],{\bf y}[(i\!-\!1)T\!+\!2],\ldots, {\bf y}[iT] \}$. By taking the sample average, it is given by 
 \begin{align}
 	{\bf \bar y}_i=\frac{1}{T}\sum_{t=(i\!-\!1)T\!+\!1}^{iT}{\bf y}[t].
 \end{align}
Since the received signal vectors $\{{\bf y}[(i\!-\!1)T\!+\!1],{\bf y}[(i\!-\!1)T\!+\!2],\ldots, {\bf y}[iT] \}$ are IID, each with a finite mean,  this centroid vector (the sample average) almost surely converges to the true mean value ${\bf \bar y}_i = \mathbb{E}[{\bf y}[t]|{\bf x}_i]=\sum_{{\bf y}_j\in \mathcal{Y}}{\bf y}_j\mathbb{P}[{\bf y}[t]={\bf y}_j|{\bf x}={\bf x}_j]$, as $T\rightarrow \infty$ by the strong law of large numbers \cite[Theorem 4.3.1]{SLLN:04}. Using ${\bf \bar y}_i$, the $i$th codebook vector is estimated as
\begin{align}
	{\bf \hat c}_i={\sf sign}({\bf \bar y}_i),
\end{align} 
where $i=\left\{1,2,\ldots, |\mathcal{M}_{\rm t}|^{K}\right\}$.
 
%  by calculating the $n$th component of the $i$th codeword vector as 
% \begin{align}
% 	{\hat c}_{i,n}=\begin{cases}-1 &\text{if}~  \sum_{t=(i-1)T+1}^{iT} y_n[t]<0\\1 &\text{if}~  \sum_{t=(i-1)T+1}^{iT}y_n[t] \geq 0\end{cases}.
% \end{align} 

 {\bf Channel parameter learning:} 
Once the codebook is constructed, we need to estimate the crossover probabilities ${\hat p}_{i,n}$ for $i\in \{1,2, \ldots, |\mathcal{M}_{\rm t}|^{K}\}$ and $n\in \{1,2,\ldots, 2N\}$ using both ${\bf Y}_t$ and $\mathcal{\hat C}=\{{\bf \hat c}_1,{\bf \hat c}_2,\ldots,{\bf \hat c}_{|\mathcal{M}_{\rm t}|^{K}}\}$. These transition probabilities are empirically estimated as
\begin{align}
 {\hat p}_{i,n}&=  \frac{1}{T}\sum_{t=(i\!-\!1)T\!+\!1}^{iT}\lVert {\hat c}_{i,n}-y_n[t]\rVert_0, % \nonumber \\
 %     {\hat p}_{i,n}^c&= 1-\frac{1}{T}\sum_{t=i+1}^{i+T}\lVert {\hat c}_{i,n}-y_n[t]\rVert_0,%\\
   % \hat{\beta}_{r,n}&=-\ln\left(1-\frac{1}{T}\sum_{t=1}^{T}\lVert \hat{c}_{r,n}-y_{n,(rT+t)}\rVert_0\right)
\end{align}
for $n\in \left\{1,...,2N\right\}$. Similarly, these estimated channel transition probabilities converge to the true distribution in probability, when the number of training samples is sufficiently large. In practical communication systems, however, the number of training samples should be small to enhance the transmission efficiency; thereby, the estimated transition probabilities and the codebook can be erroneous when using a limited number of pilots, i.e., training samples.  
%     \begin{figure*}[t]
% \center
%  \includegraphics[width=12cm]{training.pdf}
%  \caption{The proposed end-to-end supervised-learning based detection method. By sending pilots, we jointly train the model parameters including the codebook and the crossover probabilities.} \label{fig:super}
%\end{figure*} 

{\bf Detection:} Under our model assumption and the estimated model parameters, we explain an approximate-ML (A-ML) detector via the weighted minimum Hamming distance (wMHD) decoding. The A-ML detector differs from the optimal ML detector derived in \eqref{eq:ML_detection}, because it relies on our simplified 2$N$-parallel BSC model to reduce the computational complexity.

When ${\bf \hat c}_i\in \mathcal{\hat C}$ and ${\hat p}_{i,n}$ are obtained during the training phase, the log-likelihood function is given by
 \begin{align}
 \ln\mathbb{P}({\bf y}[t]|{\bf \hat c}_i)
 &=\sum_{n=1}^{2N}\ln\mathbb{P}({y}_n[t]|{\hat c}_{i,n}) \nonumber\\
 &=\sum_{n=1}^{2N}\!\left( \ln p_{i,n}{\bf 1}_{\{\hat c_{i,n}\neq y_n[t]\}}+\ln (1-p_{i,n}){\bf 1}_{\{\hat c_{i,n} = y_n[t]\}}\! \right)\!. \label{eq:model_liki}
 \end{align}
 Therefore, the A-ML detector is equivalent to the wMHD decoding as
 \begin{align}
 {\bf \hat x}_i^{\star}&=\arg\max_{{\bf \hat c}_i\in \mathcal{\hat C}}\ln\mathbb{P}({\bf y}[t]|{\bf \hat c}_i) \nonumber \\
% &=\arg\min_{{\bf \hat c}_i\in \mathcal{\hat C}}-\ln\mathbb{P}({\bf y}[t]|{\bf \hat c}_i)\nonumber \\
 %&=\arg\min_{{\bf \hat c}_i\in \mathcal{\hat C}} \sum_{n=1}^{2N} \ln\left(\frac{1}{{\hat p}_{i,n}}\right){\bf 1}_{\{{\hat c}_{i,n}\neq y_n[t]\}}\!+\!\ln \frac{1}{1\!-\!{\hat p}_{i,n}}{\bf 1}_{\{{\hat c}_{i,n} = y_n[t]\}}  \nonumber \\
 &=\arg\min_{{\bf \hat c}_i\in \mathcal{\hat C}}\! \sum_{n=1}^{2N} \!\ln\! \frac{1}{{\hat p}_{i,n}} \!|{\hat c}_{i,n}\!-\!y_n[t]|_0\!+\!\ln \!\frac{1}{1\!-\!{\hat p}_{i,n}}(\!1\!-\!|{\hat c}_{i,n}\!-\!y_n[t]|_0\!)  \nonumber \\
 &=\arg\min_{ i\in \{1,\ldots, |\mathcal{M}_{\rm t}|^{K}\}}   d_{\sf wH}\left({\bf \hat c}_i,{\bf y}[t]; \{{\hat w}_{i,n}\}_{n=1}^{2N}, \!\{{\hat w}^c_{i,n}\}_{n=1}^{2N}\!\right), \label{eq:A-ML_decoding}
 \end{align}
where the last equality comes from the definition of the weighted Hamming distance with the weights ${\hat w}_{i,n}=-\ln \left({\hat p}_{i,n}\right)$ and ${\hat w}^c_{i,n}=-\ln \left(1- {\hat p}_{i,n}\right)$ in \cite{Hong:18}.  The proposed supervised-learning communication framework is summarized in Algorithm \ref{alg:SL_model}.

\begin{algorithm}
	\caption{The proposed end-to-end supervised-learning framework.}\label{alg:SL_model}
	{\small{\begin{algorithmic}[1]
				\FOR [Training for parameter learning] {$t=1,\ldots, T_{\rm t}$}	
				\STATE Centroid update ${\bf \bar y}_i=\frac{1}{T}\sum_{t=(i\!-\!1)T\!+\!1}^{iT}{\bf y}[t]$.
			\ENDFOR
			\STATE Compute the codeword vectors ${\bf \hat c}_i={\sf sign}({\bf \bar y}_i)$ for $i=\left\{1,2,\ldots, |\mathcal{M}_{\rm t}|^{K}\right\}$.
			\STATE Compute the transition probabilities $ {\hat p}_{i,n}=  \frac{1}{T}\sum_{t=(i\!-\!1)T\!+\!1}^{iT}\lVert {\hat c}_{i,n}-y_n[t]\rVert_0$.
	\FOR[Data detection with the trained parameters] {$t=T_{\rm t}+1,\ldots, T_{\rm B}$}	
				\STATE Perform the A-ML detection: ${\bf \hat x}_i^{\star}= \arg\min_{i}   d_{\sf wH}\left({\bf \hat c}_i,{\bf y}[t]; \{{\hat w}_{i,n}\}_{n=1}^{2N}, \!\{{\hat w}^c_{i,n}\}_{n=1}^{2N}\!\right)$\\ with ${\hat w}_{i,n}=-\ln \left({\hat p}_{i,n}\right)$ and ${\hat w}^c_{i,n}=-\ln \left(1- {\hat p}_{i,n}\right)$.
			\ENDFOR
	\end{algorithmic}}}
\end{algorithm}

%{\bf Remark 6:} Our proposed the approximated-ML detector does not consider all possible received signals at the relays. Whereas, the ML detector considers the all possible received signals at each layer. Therefore, the performance loss can occur when using the approximated-ML detector in \eqref{eq:A-ML_decoding}, compared to the ML detector in \eqref{eq:ML_detection}.

 {\bf Remark 4 (Model parameter reduction):}  The most advantageous feature of the proposed one is the huge reduction of the number of model parameters to perform data detection. To make a quantitative claim for this, we compare the number of parameters required for the data detection in the two channel models. In the original channel, a total number of model (real-valued) parameters needed for the classical ML detector is $4KL + (M-2)(4L^2) + 4LN+(M-2)L+1$, which includes the number of multi-hop channel elements and SNRs per hop. As can be seen, the number of model parameters scales linearly with $K$, $N$, and $M$, while it quadratically increases with $L$. Whereas, the proposed model for the supervised-learning requires much fewer model parameters. Since it only considers the transition probabilities of each channel input, a total of $|\mathcal{M}_{\rm t}|^{K}2N$ model parameters is required to perform the A-ML detector. Although the number of model parameters scales exponentially with $K$, it does not scale with the number of hops $M$ and the number of distributed relays per layer $L$. Therefore, when the number of co-scheduled uplink users $K$ is a few, the proposed model can reduce the model parameters significantly.

{\bf Remark 5 (Detection complexity reduction):} The proposed A-ML detector does not require perfect and global CSI knowledge at the BS. Instead, the empirically estimated codebook $ \mathcal{\hat C}$ and the channel weights $\{{\hat w}_{i,n}\}_{n=1}^{2N}$ are sufficient, which can be accurately estimated from the simple repetition training strategy, even using a reasonable amount of pilots. In addition, the computational complexity of this method is the order of $\mathcal{O}(|\mathcal{M}_{\rm t}|^{K})$, which does not scale with the number of relays and hops. This is a huge complexity reduction compared to the original ML detector in Section III. The required training length, however, exponentially increases with the number of uplink users as in the single-hop multi-user MIMO system with one-bit ADCs. Therefore, for the implementation, the number of co-scheduled uplink users should be chosen to meet the constraint of pilot overheads or a semi-supervised-learning and reinforcement-learning methods can be used as in \cite{Kim:19} and \cite{Jeon:TWC:19}. In addition, the complexity of the A-ML detector can be further reduced using one-bit sphere decoding in \cite{Jeon:TWC:18}. 

 \subsection{ Optimality of the Proposed Model}
The proposed end-to-end network model simplifies the classical multi-hop channel model by significantly reducing the number of model parameters. This simplification can cause the model errors in general.  In this subsection, we show that the proposed end-to-end network model can still be optimal in terms of the detection performance under a certain scenario. 
%In this subsection, we prove that the proposed simple end-to-end network model guarantees the optimality in the sense of minimizing the symbol error probability in certain conditions.

%{\bf  Remark 5 (The optimality of the proposed model and detection method):} The proposed parallel BSC model can be optimal when the channel matrices of the first $M-1$ hops are diagonal. For instance, when considering a mmWave wireless backhaul relaying system illustrated in Fig. \ref{Fig2}, the first-hop channel matrix can be modeled as a diagonal matrix, because the inter-user interference at the relays can be eliminated by analog beamforming. In this case, the detection performance of the proposed A-ML is equivalent to that of the ML detection presented in Section III, even with a reasonable amount of pilots.  This effect will be verified in Section VII. 
{\bf Theorem 1:} The proposed parameter learning and detection method is optimal in the sense of minimizing the detection error when i) $\sigma_1=\cdots=\sigma_{M-1}\rightarrow 0$ and ii) the training length per channel input, $T$, is large enough.

To prove Theorem 1, we need the following lemma, which elucidates that the proposed simple training method guarantees the optimality for the parameter learning, provided that  the number of training samples is sufficiently large and the SNRs of the $M-1$ hops are infinite.  %Although this is a special case, it elucidates why the proposed method performs well. 

{\bf Lemma 1:} Let $\frac{1}{T}\sum_{t=(i-1)T+1}^{iT}{\bf y}[t]$ be the received sample average when ${\bf x}_i\in \mathcal{X}$ was sent during the training phase. Under the infinite SNR assumptions of the $M-1$ hops, i.e., $\sigma_i=0$ for $i=1,2,\ldots, M-1$, the sign of the sample average for the received vectors almost surely converges to the codeword vector ${\bf c}_i={\sf sign}\left({\bf H}_{M}\cdots{\sf sign}\left({\bf H}_1{\bf  x}_i\right)\right)$ as $T\rightarrow \infty$, i.e., 
\begin{align}
	{\sf sign}\left(\lim_{T\rightarrow \infty} \frac{1}{T}\sum_{t=(i-1)T+1}^{iT}{\bf y}[t]\right) = {\bf c}_i.
\end{align}
In addition, the empirical transition probability converges to the true one as as $T\rightarrow \infty$, namely,
\begin{align}
	 \lim_{T\rightarrow \infty} \frac{1}{T}\sum_{t=(i\!-\!1)T\!+\!1}^{iT}\lVert {c}_{i,n}-y_n[t]\rVert_0 =Q\left(\frac{|g_{i,n}|}{\sigma_M/\sqrt{2}}\right),
\end{align}
where $g_{i,n}$ is the $n$th element of ${\bf g}_i={\bf H}_{M}\cdots{\sf sign}\left({\bf H}_1{\bf  x}_i\right) \in \mathbb{R}^{2N}$.

\begin{proof}
Under the premise that $\sigma_i=0$ for $i=1,2,\ldots, M-1$, the received signal at time slot $t$ is rewritten as
\begin{align}
	{\bf y}[t] &=  {\sf sign}\left({\bf H}_{M}\cdots{\sf sign}\left({\bf H}_1{\bf  x}_i\right)+{\bf v}_M[t]\right) \nonumber\\
	&={\sf sign}\left({\bf g}_i+{\bf v}_M[t]\right).
\end{align}
Since ${\bf v}_M[t]$ is IID over $t$, by the law of large numbers, the sample average converges to its mean, namely,
\begin{align}
	\lim_{T\rightarrow \infty} \frac{1}{T}\sum_{t=(i-1)T+1}^{iT}{\bf y}[t] &=\mathbb{E}\left[{\bf y}[t]|{\bf g}_i\right].\end{align}
Let ${\bf \bar y}_i=\mathbb{E}\left[{\bf y}[t]|{\bf g}_i\right]$. Then, the $n$th component of ${\bf \bar y}_i$ is computed as
\begin{align}
	{\bar y}_{i,n} &= \mathbb{P}\left[ {\sf sign}\left(g_{i,n}+ v_{M,n}[t]\right) = {\sf sign}\left(g_{i,n}\right)\right]{\sf sign}\left(g_{i,n}\right)  + \mathbb{P}\left[ {\sf sign}\left(g_{i,n}+ v_{M,n}[t]\right) \neq {\sf sign}\left(g_{i,n}\right)\right]\left(-{\sf sign}\left(g_{i,n}\right)\right) \nonumber\\ 
	&=\left( 1-2 \mathbb{P}\left[ {\sf sign}\left(g_{i,n}+ v_{M,n}[t]\right) \neq {\sf sign}\left(g_{i,n}\right)\right]\right) {\sf sign}\left(g_{i,n}\right)
	\nonumber\\ 
	&=\left( 1-2 Q\left(\frac{|g_{i,n}|}{\sigma_M/\sqrt{2}}\right)\right) {\sf sign}\left(g_{i,n}\right).
	\end{align}
	Since $Q\left(\frac{|g_{i,n}|}{\sigma_M/\sqrt{2}}\right)<0.5$ for $\frac{|g_{i,n}|}{\sigma_M/\sqrt{2}}>0$, $\left( 1-2 Q\left(\frac{|g_{i,n}|}{\sigma_M/\sqrt{2}}\right)\right)$ does not change the sign. Therefore, we conclude that 
	\begin{align}
		{\sf sign}({\bar y}_{i,n} ) &= {\sf sign}(g_{i,n})={c}_{i,n}, \label{eq:codeword_convergence}
	\end{align}
	for all $n\in \{1,2,\ldots, 2N\}$. Further, when the number of the training samples is large enough, $T\rightarrow \infty$, the empirical transition probability ${\hat p}_{i,n}$ converges to
	\begin{align}
\lim_{T\rightarrow \infty} {\hat p}_{i,n}&= \lim_{T\rightarrow \infty}\frac{1}{T}\sum_{t=(i\!-\!1)T\!+\!1}^{iT}\lVert {\hat c}_{i,n}-y_n[t]\rVert_0  \nonumber\\
&\stackrel{(a)}{=} \lim_{T\rightarrow \infty}\frac{1}{T} \sum_{t=(i\!-\!1)T\!+\!1}^{iT}\lVert { c}_{i,n}-y_n[t]\rVert_0  \nonumber\\
&\stackrel{(b)}{=} \lim_{T\rightarrow \infty}\frac{1}{T} \sum_{t=(i\!-\!1)T\!+\!1}^{iT} {\bf 1}_{\left\{ {\sf sign}\left(g_{i,n}\right) \neq  {\sf sign}\left(g_{i,n}+ v_{M,n}[t]\right) \right\}} \nonumber\\
&\stackrel{(c)}{=} \mathbb{E}\left[{\bf 1}_{\left\{ {\sf sign}\left(g_{i,n}\right) \neq  {\sf sign}\left(g_{i,n}+ v_{M,n}[t]\right) \right\}}\right]
  \nonumber\\
 &= \mathbb{P}\left[ {\sf sign}\left(g_{i,n}\right) \neq  {\sf sign}\left(g_{i,n}+ v_{M,n}[t]\right) \right] \nonumber\\
 &=Q\left(\frac{|g_{i,n}|}{\sigma_M/\sqrt{2}}\right),
\end{align}
where (a) follows from \eqref{eq:codeword_convergence}, (b) holds by the definitions of the indicator function, $c_{i,n}={\sf sign}(g_{i,n})$, and $y_n[t]={\sf sign}\left(g_{i,n}+ v_{M,n}[t]\right)$, and (c) is by the law of large numbers.  
\end{proof}

Now, we are ready to prove Theorem 1.

\begin{proof}
From Lemma 1, when $T$ is sufficiently large, the model parameters including the codebook and the effective transition probabilities can be perfectly estimated. Therefore, to prove Theorem 1,  it is sufficient to show that the transition probability in \eqref{eq:likelihood} is equivalent to $\mathbb{P}({\bf y}_j|{\bf c}_i)$ under the assumption of $\sigma_1=\cdots=\sigma_{M-1}\rightarrow 0$. To do this, from the assumption, we first simplify the probability that the BS receives ${\bf y}_j\in \mathcal{Y}$ when ${\bf x}_i\in \mathcal{X}$ was sent in \eqref{eq:likelihood} as
\begin{align}
	\lim_{\sigma_1,\ldots,\sigma_{M-1}\rightarrow 0}\mathbb{P}\left({\bf y}={\bf y}_{j}|{\bf x}={\bf x}_i\right)&= \mathbb{P}\left( {\bf y}_j =  {\sf sign}\left({\bf H}_{M}\cdots{\sf sign}\left({\bf H}_1{\bf  x}_i\right)+{\bf v}_M\right)   \right) \nonumber\\
	&=\mathbb{P}\left( {\bf y}_j = {\sf sign}\left({\bf g}_i +{\bf v}_M\right)  \right) \nonumber\\
	&=\prod_{n=1}^{2N}\mathbb{P}(y_{j,n}= {\sf sign}\left(g_{i,n}+v_{M,n}\right)) \nonumber\\
	&=\prod_{n=1}^{2N}   Q\left(\frac{|g_{i,n}|}{\mathbf{\sigma}_{M}/ \sqrt{2}}\right )^{{\bf 1}_{ \left\{y_{j,n}\neq {\sf sign}(g_{i,n})\right\}}} \times Q\left(\frac{ -|g_{i,n}|}{\mathbf{\sigma}_{M}/ \sqrt{2}}\right )^{{\bf 1}_{ \left\{y_{j,n}={\sf sign}(g_{i,n})\right\}}}.%  \nonumber\\
	%&=\prod_{n=1}^{2N} Q\left(\frac{- y_{j,n}c_{i,n}}{\mathbf{\sigma}_{M}/ \sqrt{2}}\right ).
\end{align}
By taking a log function, the corresponding log-likelihood function of the nonlinear multi-hop channel becomes
\begin{align}
	\lim_{\sigma_1,\ldots,\sigma_{M-1}\rightarrow 0}\ln\mathbb{P}\left({\bf y}={\bf y}_{j}|{\bf x}={\bf x}_i\right)= \sum_{n=1}^{2N}\!\left( \ln p_{i,n}{\bf 1}_{\{ c_{i,n}\neq y_{j,n}\}}+\ln (1-p_{i,n}){\bf 1}_{\{ c_{i,n} = y_{j,n}\}}\! \right), 
\end{align}
where $p_{i,n}= Q\left(\frac{y_{j,n}c_{i,n}}{\mathbf{\sigma}_{M}/ \sqrt{2}}\right)$, which completes the proof.
\end{proof}

\color{black}\section{Model-based Online Supervised-Learning Approach}

In this section, we propose an end-to-end online supervised-learning detector, which is robust to time-varying channel environments. We explain the proposed online-learning detector using the expectation-maximization framework.  The proposed detector iteratively finds maximum likelihood estimates of the model parameter using the received signals during the data transmission phase by treating them as new training (labelled) data samples.  

%, during the data transmission phase, the proposed online-learning algorithm jointly updates the model parameters (the codebook and the crossover probabilities) by exploiting reliably detected data symbols.

%The scenario of the proposed communication framework consists of channel training and data detection. In the channel training phase, the receiver learns the initial centroid of received symbol vector set. A pilot symbol sequence is composed by the least number of symbol vectors to resolve the permutation ambiguity. It consists of all possible symbol vectors given as 

%\subsection{EM-based Parameter Learning}

%The log-likelihood function is 
%
%\begin{align}
%	\ln\mathbb{P}\left({\bf y}={\bf y}_{j}|{\bf x}={\bf x}_i\right)= \sum_{n=1}^{2N}\!\left( \ln p_{i,n}{\bf 1}_{\{ c_{i,n}\neq y_{j,n}\}}+\ln (1-p_{i,n}){\bf 1}_{\{ c_{i,n} = y_{j,n}\}}\! \right), 
%\end{align}
%where $p_{i,n}=\ln Q\left(\frac{y_{j,n}c_{i,n}}{\mathbf{\sigma}_{M}/ \sqrt{2}}\right)$.

 \subsection{Joint Probability Distribution for Labeled and Unlabelled Training Samples }

We denote a set of the model parameters of the end-to-end network by $\Theta= \left\{ \{{ p}_{i,n}, c_{i,n}\}_{n=1}^{2N}\right\}_{i=1}^{|\mathcal{M}_{\rm t}|^{K}}$. We also define a binary vector ${\bf z}[t]=[z_1[t], z_2[t],\ldots, z_{|\mathcal{M}_{\rm t}|^{K}}[t]]$, where $z_i[t]\in \{1,0\}$ indicates whether the $i$th information vector was sent or not at time $t$. Therefore, $\sum_{i=1}^{|\mathcal{M}_{\rm t}|^{K}}z_i[t]=1$.  For the training phase, i.e., $t\in \{1,2,\ldots, T_{\rm t}\}$, ${\bf z}[t]$ is a deterministic vector because the training samples are labeled. For instance, for $t\in \mathcal{T}_i$, $z_i[t]=1$ and $z_{k}[t]=0$ for $k\neq i$. Whereas, for the data transmission phase $t\in \{T_{\rm t}+1,\ldots, T_{\rm B}\}$, ${\bf z}[t]$ is a hidden variable, i.e., a random vector, because  the received signal vector ${\bf y}[t]$ is unlabeled.  Our goal is to design an algorithm that jointly perform the data detection (e.g., the assignment of a label) and the update of the model parameters by harnessing both the labeled received signals ${\bf y}[t]$ for $t\in \cup_{i=1}^{|\mathcal{M}_{\rm t}|^{K}}\mathcal{T}_i$ and the unlabeled received signal vectors ${\bf y}[t]$ for $t\in \mathcal{T}_{\tau}=\{T_{\rm t}+1,\ldots, \tau\}$ where $T_{\rm t}+1\leq \tau\leq T_{\rm B}$. To accomplish this, we propose an online learning algorithm inspired by the expectation-maximization framework.  

 Using the proposed model developed in Section IV, we define a joint probability distribution for $\{ {\bf y}[t]\}_{t=1}^{\tau}$ conditioned that the model parameter set $\Theta$ and the labels ${\bf z}[t]$ are given as
\begin{align}
	p\left(\{ {\bf y}[t]\}_{t=1}^{\tau},\mid  \{{\bf z}[t]\}_{t=1}^{\tau},\Theta \right) \!=\!\prod_{n=1}^{2N}\prod_{i=1}^{|\mathcal{M}_{\rm t}|^{K}}\prod_{t=1}^{\tau}\!\left[ {p_{i,n}}^{{\bf 1}_{\{ c_{i,n}  \neq y_{n}[t]\}}}\left(1\!-\!p_{i,n}\right)^{{\bf 1}_{\{ c_{i,n} = y_{n}[t]\}}}\!\right]^{\!z_{i}[t]}.
\end{align}
We also define the joint probability of the labels as 
\begin{align}
	p\left(  \{{\bf z}[t]\}_{t=1}^{\tau} \right) =\prod_{i=1}^{|\mathcal{M}_{\rm t}|^{K}}\prod_{t=1}^{\tau}\pi_i^{z_{i}[t]},
\end{align}
where $\pi_i=\mathbb{P}[{\bf x}={\bf x}_i]=\frac{1}{|\mathcal{M}_{\rm t}|^{K}}$. Then, by Bayes' rule,  the joint probability of $\{ {\bf y}[t]\}_{t=1}^{\tau}$ and $\{{\bf z}[t]\}_{t=1}^{\tau}$ when the model parameter set $\Theta$ is given can be written as 
\begin{align}
	p\left(\{ {\bf y}[t]\}_{t=1}^{\tau},\{{\bf z}[t]\}_{t=1}^{\tau}\mid  \Theta \right) &=p\left(\{ {\bf y}[t]\}_{t=1}^{\tau}\mid  \{{\bf z}[t]\}_{t=1}^{\tau},\Theta \right)p\left(  \{{\bf z}[t]\}_{t=1}^{\tau} \right) \nonumber \\	
	&\!\!\!\!\!\!=\!\prod_{n=1}^{2N}\prod_{i=1}^{|\mathcal{M}_{\rm t}|^{K}}\!\prod_{t=1}^{\tau}\left[ {p_{i,n}}^{{\bf 1}_{\{ c_{i,n} \neq y_{n}[t]\}}}\left(1\!-\!p_{i,n}\right)^{{\bf 1}_{\{ c_{i,n} \!=\! y_{n}[t]\}}}\!\right]^{\!z_{i}[t]}\prod_{t=1}^{\tau}\prod_{i=1}^{|\mathcal{M}_{\rm t}|^{K}}\pi_i^{z_{i}[t]}.
\end{align}
Then, the log-likelihood function of $({\bf y}[t]\}_{t=1}^{\tau},\{{\bf z}[t]\}_{t=1}^{\tau})$ conditioned on $\Theta$ is given by 
\begin{align}
	&\ln p\left(\{ {\bf y}[t]\}_{t=1}^{\tau},\{{\bf z}[t]\}_{t=1}^{\tau}\mid  \Theta \right) \nonumber\\
	&=\!\sum_{n=1}^{2N}\sum_{i=1}^{|\mathcal{M}_{\rm t}|^{K}}\sum_{t=1}^{\tau}\!z_{i}[t]\left[ \ln{p_{i,n}}{\bf 1}_{\{ c_{i,n}  \neq y_{n}[t]\}}+ \ln\left(1\!-\!p_{i,n} \right){\bf 1}_{\{ c_{i,n} \!=\! y_{n}[t]\}}\!\right] +\sum_{t=1}^{\tau}\sum_{i=1}^{|\mathcal{M}_{\rm t}|^{K}}z_{i}[t] \ln \pi_i. \label{eq:log_liklihood}
\end{align}

\subsection{Expectation Step}
In the expectation step, we compute the probability that the $k$th label is assigned to a new received signal vector. Recall that ${\bf z}[t]$ is fixed for $t=\{1,2,\ldots, T_{\rm t}\}$. Therefore, in this step, we estimate the hidden variable ${\bf z}[t]$  for $t\in \mathcal{T}_{\tau}=\{T_{\rm t}+1,\ldots, \tau\}$ by taking the expectation to the log-likelihood function in \eqref{eq:log_liklihood} with respective to the conditional distribution $p\left({\bf z}[t] \mid \{ {\bf y}[t]\}_{t=1}^{\tau}, \Theta\right)$, namely, 
\begin{align}
	&\mathbb{E}\left[\ln p\left(\{ {\bf y}[t]\}_{t=1}^{\tau},\{{\bf z}[t]\}_{t=1}^{\tau}\mid  \Theta \right)\right] \nonumber\\
	&=\!\sum_{n=1}^{2N}\sum_{i=1}^{|\mathcal{M}_{\rm t}|^{K}}\sum_{t=1}^{\tau}\!\mathbb{E}\left[z_{i}[t] \mid \{ {\bf y}[t]\}_{t=1}^{\tau}, \Theta\right]\left[ \ln{p_{i,n}}{\bf 1}_{\{ c_{i,n} \neq y_{n}[t]\}}+ \ln\left(1\!-\!p_{i,n}\right){\bf 1}_{\{ c_{i,n}\!=\! y_{n}[t]\}}\!\right] \nonumber\\
	&+\sum_{i=1}^{|\mathcal{M}_{\rm t}|^{K}}\sum_{t=1}^{\tau}\mathbb{E}\left[z_{i}[t] \mid \{ {\bf y}[t]\}_{t=1}^{\tau}, \Theta\right] \ln \pi_i \nonumber\\
	&=\!\sum_{n=1}^{2N}\sum_{i=1}^{|\mathcal{M}_{\rm t}|^{K}}\!\mathbb{P}\left(z_{i}[t]=1 \mid \{ {\bf y}[t]\}_{t=1}^{\tau}, \Theta\right)\left[ \ln{p_{i,n}}{\bf 1}_{\{ c_{i,n} \neq y_{n}[t]\}}+ \ln\left(1\!-\!p_{i,n}\right){\bf 1}_{\{ c_{i,n} \!=\! y_{n}[t]\}}\!\right] \nonumber \\
	&+\sum_{i=1}^{|\mathcal{M}_{\rm t}|^{K}}\sum_{t=1}^{\tau}\mathbb{P}\left(z_{i}[t]=1 \mid \{ {\bf y}[t]\}_{t=1}^{\tau}, \Theta\right) \ln \pi_i,
\end{align}
where the last equality follows from the fact that $z_i[t]$ is an indicator function, i.e., $\mathbb{E}\left[z_{i}[t]\mid \{ {\bf y}[t]\}_{t=1}^{\tau}, \Theta\right]=\mathbb{P}\left(z_{i}[t]=1\mid \{ {\bf y}[t]\}_{t=1}^{\tau}, \Theta\right)$. Recall that $\mathbb{P}[z_i[t]=1\mid \{ {\bf y}[t]\}_{t=1}^{\tau}, \Theta]=1$ for $t\in \mathcal{T}_i$ where $i\in \{1,2,\ldots, |\mathcal{M}_{\rm t}|^{K}\}$ thanks to the training phase. The probability that the $k$th label is assigned to the new observation ${\bf y}[t]$ for $t\in \mathcal{T}_{\tau}=\{T_{\rm t}+1,\ldots, \tau\}$ is obtained by computing  APP as
\begin{align}
	\mathbb{P}[z_{k}[\tau]\!=\!1\mid \{ {\bf y}[t]\}_{t=1}^{\tau}, \Theta]
		&=\frac{\mathbb{P}[z_{k}[\tau]=1, {\bf y}[\tau]\ \mid  \Theta] \mathbb{P}[z_k[\tau]=1]}{\mathbb{P}[ {\bf y}[\tau] \mid  \Theta]}\nonumber\\
	&=\frac{\prod_{n=1}^{2N} \left[ {p_{k,n}}^{{\bf 1}_{\{ c_{k,n} \neq y_{n}[\tau]\}}}\left(1\!-\!p_{k,n}\right)^{{\bf 1}_{\{ c_{k,n} \!=\! y_{n}[t]\}}}\!\right]  \pi_k}{ \sum_{i=1}^{|\mathcal{M}_{\rm t}|^{K}} \pi_i \prod_{n=1}^{2N}  \left[ {p_{i,n}}^{{\bf 1}_{\{ c_{i,n} \neq y_{n}[\tau]\}}}\left(1\!-\!p_{i,n}\right)^{{\bf 1}_{\{ c_{i,n} \!=\! y_{n}[t]\}}}\!\right]     }\nonumber\\
	&=\frac{\prod_{n=1}^{2N} \left[ {p_{k,n}}^{{\bf 1}_{\{ c_{k,n} \neq y_{n}[t]\}}}\left(1\!-\!p_{k,n}\right)^{{\bf 1}_{\{ c_{k,n} \!=\! y_{n}[t]\}}}\!\right]  }{ \sum_{i=1}^{|\mathcal{M}_{\rm t}|^{K}}\prod_{n=1}^{2N} \left[ {p_{i,n}}^{{\bf 1}_{\{ c_{i,n} \neq y_{n}[\tau]\}}}\left(1\!-\!p_{i,n}\right)^{{\bf 1}_{\{ c_{i,n} \!=\! y_{n}[\tau]\}}}\!\right]  }, \label{eq:EM_APP}
\end{align}
where the last equality follows from $\pi_i=\pi_k$ for all $i$ and $k$. From \eqref{eq:model_liki}, since ${p_{i,n}}^{{\bf 1}_{\{ c_{i,n} \neq y_{n}[t]\}}}\left(1\!-\!p_{i,n}\right)^{{\bf 1}_{\{ c_{i,n} \!=\! y_{n}[t]\}}}\!=e^{ -d_{\sf wH}\left({\bf c}_i,{\bf y}[\tau]; \{\ln (p_{i,n})\}_{n=1}^{2N},\{\ln (1-p_{i,n})\}_{n=1}^{2N}\right)}$, we rewrite the APP in \eqref{eq:EM_APP} as
\begin{align}
	\gamma_k[\tau] &= \mathbb{P}[z_{k}[\tau]\!=\!1\mid  {\bf y}[\tau], \Theta] \nonumber\\	&= \frac{\exp\left(-d_{\sf wH}\left({\bf c}_{k},{\bf y}[\tau] ;\{{w}_{k,n}\}_{n=1}^{2N},\{{w}^c_{k,n}\}_{n=1}^{2N}\right)\right)}{\sum_{j=1}^{|\mathcal{M}_{\rm t}|^{K}}\exp\left(-d_{\sf wH}\left({\bf c}_j,{\bf y}[\tau]; \{{w}_{j,n}\}_{n=1}^{2N},\{{ w}^c_{j,n}\}_{n=1}^{2N}\right)\right)}.\label{eq:APP}
\end{align}
Therefore, to compute the reliability of the label $\gamma_k[\tau]$ for ${\bf y}[\tau]$, we need the estimated model parameters using the previously received signal vectors $\{ {\bf y}[t]\}_{t=1}^{\tau-1}$. Let ${\hat \Theta}[\tau\!-\!1]=\{ {\hat p}_{i,n}[\tau\!-\!1], {\hat c}_{i,n}[\tau\!-\!1]\}_{n=1}^{2N}\}$ be the estimated model parameter set using the received signal vectors $\{ {\bf y}[t]\}_{t=1}^{\tau-1}$. Then, the APP in \eqref{eq:APP} can be rewritten in terms of ${\hat \Theta}[\tau\!-\!1]$ as
\begin{align}
	{\hat \gamma}_k[\tau]= \frac{\exp\left(-d_{\sf wH}\left({\bf \hat c}_{k}[\tau\!-\!1],{\bf y}[\tau] ;\{{\hat w}_{k,n}[\tau\!-\!1]\}_{n=1}^{2N},\{{\hat w}^c_{k,n}[\tau\!-\!1]\}_{n=1}^{2N}\right)\right)}{\sum_{j=1}^{|\mathcal{M}_{\rm t}|^{K}}\exp\left(-d_{\sf wH}\left({\bf c}_j[\tau\!-\!1],{\bf y}[\tau]; \{{\hat w}_{j,n}[\tau\!-\!1]\}_{n=1}^{2N},\{{\hat w}^c_{j,n}[\tau\!-\!1]\}_{n=1}^{2N}\right)\right)}.\label{eq:APP_esti}
\end{align}

\subsection{Maximization Step}
In the maximization step, we need to find the parameters that maximize the expected log-likelihood function. Since the term $\sum_{i=1}^{|\mathcal{M}_{\rm t}|^{K}}\sum_{t=1}^{\tau}{\hat \gamma}_i[t] \ln \pi_i$ is irrelevant to the parameter estimation, an effective log-likelihood function for the parameter optimization is 
\begin{align}
	\mathcal{L}(\Theta)=&\mathbb{E}\left[\ln p\left(\{ {\bf y}[t]\}_{t=1}^{\tau},\{{\bf z}[t]\}_{t=1}^{\tau}\mid  \Theta \right)\right]-\sum_{i=1}^{|\mathcal{M}_{\rm t}|^{K}}\sum_{t=1}^{\tau}{\hat \gamma}_i[t] \ln \pi_i\nonumber\\
	&=\!\sum_{n=1}^{2N}\sum_{i=1}^{|\mathcal{M}_{\rm t}|^{K}}\sum_{t=1}^{\tau}{\hat \gamma}_i[t]\left[ \ln{p_{i,n}}{\bf 1}_{\{ c_{i,n} \neq y_{n}[t]\}}+ \ln\left(1\!-\!p_{i,n}\right){\bf 1}_{\{ c_{i,n} \!=\! y_{n}[t]\}}\!\right].\label{eq:expected_logliki}
\end{align}
Unfortunately, maximizing $\mathcal{L}(\Theta)$ with respective to $\Theta$ is  a mixed integer optimization problem, because ${\bf c}_i$ is a discrete vector. Instead of jointly finding the parameters, we solve this optimization problem with a two-step approach: 1) the codeword estimation and 2) the transition probability estimation.  

We first present the codeword estimation method. With the knowledge of $p_{i,n}$, the $n$th element of the $i$th codeword vector, $c_{i,n}\in \{-1,1\}$ is obtained by solving the following optimization problem: 
\begin{align}
	c_{i, n}^{\star}	&=\arg\max_{c_{i,n}\in\{1,-1\}}\sum_{t=1}^{\tau}{\hat \gamma}_i[t]\left[ \ln{p_{i,n}}{\bf 1}_{\{ c_{i,n} \neq y_{n}[t]\}}+ \ln\left(1\!-\!p_{i,n}\right){\bf 1}_{\{ c_{i,n} \!=\! y_{n}[t]\}}\!\right].
\end{align}
For given $p_{i,n}$, the optimal solution for $c_{i,n}^{\star}$ can be expressed using the sign function as
\begin{align}
	%c_{i,n}^{\star} &={\bf 1}_{ \{\kappa_1 > \kappa_{-1}\}} - {\bf 1}_{ \{\kappa_1 < \kappa_{-1}\}} \nonumber \\
	c_{i,n}^{\star} &={\sf sign}(\kappa_1 - \kappa_{-1}), \label{eq:c_estimate}
\end{align}
where  $	\kappa_1 = \sum_{t=1}^{\tau}{\hat \gamma}_i[t]\left[ \ln{p_{i,n}}{\bf 1}_{\{   y_{n}[t]\neq 1\}}+ \ln\left(1\!-\!p_{i,n}\right){\bf 1}_{\{ y_{n}[t]=1\}}\!\right]$ and\\ $
	\kappa_{-1}= \sum_{t=1}^{\tau}{\hat \gamma}_i[t]\left[ \ln{p_{i,n}}{\bf 1}_{\{ y_{n}[t]\neq -1\}}+ \ln\left(1\!-\!p_{i,n}\right){\bf 1}_{\{ y_{n}[t]=-1\}}\!\right]$. 
	The estimator in \eqref{eq:c_estimate} cannot be used in practice when the knowledge of $p_{i,n}$ is absent. To resolve this problem, we propose a simple blind estimator of $c_{i,n}$ using $\{y_{n}[t]\}_{t=1}^{\tau}$. Assuming that the SNR per hop is sufficiently large, i.e., $p_{i,n}\rightarrow 0$, the optimal estimator can be approximated as
\begin{align}
{\hat c}_{i,n}[\tau] &=\lim_{p_{i,n}\rightarrow 0}{\sf sign}(\kappa_1 - \kappa_{-1}) \nonumber\\
	&={\sf sign}\!\left(\!\ln{p_{i,n}}\!\left\{ \sum_{t=1}^{\tau}\gamma_i[t] \left\{ {\bf 1}_{\{ y_{n}[t] = -1\} } \!-\! {\bf 1}_{\{ y_{n}[t] = 1\} }\! \right\}\!\right\}\right)\nonumber\\
	& ={\sf sign}\left(  \sum_{t=1}^{\tau}\gamma_i[t] \left\{ {\bf 1}_{\{ y_{n}[t] = 1\} } - {\bf 1}_{\{ y_{n}[t] = -1\} }  \right\}\right)\nonumber\\
	& ={\sf sign}\left(  \sum_{t=1}^{\tau}\gamma_i[t] y_n[t]  \right),
\end{align}
where the second equality follows from the fact $\ln{p_{i,n}}<0$ for any $0< p_{i,n}< 0.5$ and the last equality is because ${\bf 1}_{\{ y_{n}[t] = 1\} } - {\bf 1}_{\{ y_{n}[t] = -1\}} =y_n[t]$. As can be seen, this estimator is not only simple to compute, but also it does not require the knowledge of $p_{i,n}$. Nevertheless, this simple weighted sample average estimator guarantees the optimality in a certain condition as shown in Lemma 1. 

%we propose a simple weighted sample average estimator to estimate ${\hat c}_{i,n}$, i.e., 
%\begin{align}
%	{\hat c}_{i,n} &= {\sf sign}\left(\frac{\sum_{t=1}^{\tau}\gamma_i[t]y_n[t]}{\sum_{t=1}^{\tau} \gamma_i[t]}\right) \nonumber \\
%	 &= {\sf sign}\left(\frac{\sum_{t\in T_i}y_n[t]+\gamma_i[\tau]y[\tau]}{T+\gamma_i[\tau]}\right).
%\end{align}
For given ${\hat c}_{i,n}[\tau]$, the expected log-likelihood function in \eqref{eq:expected_logliki} is a concave function with respective to $p_{i,n}$. Therefore, the optimal $p_{i,n}$ is obtained by taking the first-order derivative of $\mathcal{L}( p_{i,n}, {\hat c}_{i,n}[\tau] )$ with respective to $p_{i,n}$, which yields
\begin{align}
	\frac{\mathcal{L}( p_{i,n}, {\hat c}_{i,n}[\tau] )}{\partial p_{i,n}} =\sum_{t=1}^{\tau}{\hat \gamma}_i[t] \left\{\frac{1}{p_{i,n}}{\bf 1}_{\{ {\hat c}_{i,n}[\tau] \neq y_{n}[t]\}}- \frac{1}{1-p_{i,n}}{\bf 1}_{\{ {\hat c}_{i,n}[\tau]  = y_{n}[t]\}}\right\}.
\end{align}
By solving $\frac{\mathcal{L}( p_{i,n}, {\hat c}_{i}[\tau])}{\partial p_{i,n}}=0$, we obtain the optimal estimate of the $n$th transition probability when the $i$th codeword was sent as
\begin{align}
	{\hat p}_{i,n}^{\star}[\tau] &= \frac{\sum_{t=1}^{\tau}{\hat \gamma}_i[t]{\bf 1}_{\{ {\hat c}_{i,n}[\tau] \neq y_{n}[t]\}}}{\sum_{t=1}^{\tau} {\hat \gamma}_i[t]}. %\nonumber \\
%&=\frac{\sum_{t\in\mathcal{T}_i}{\bf 1}_{\{ {\hat c}_{i,n}[\tau] \neq y_{n}[t]\}} +{\hat \gamma}_i[\tau] {\bf 1}_{\{ {\hat c}_{i,n} \neq y_{n}[t]\}}}{T+\sum_{t=T_{\rm t}+1}^{\tau}{\hat \gamma}_i[t]}. 
\end{align}
Therefore, the estimated model parameter set ${\hat \Theta}[\tau]$ with the received signal vectors $\{{\bf y}[t]\}_{t=1}^U$ is 
\begin{align}
	{\hat \Theta}[\tau]=\left({\hat p}_{i,n}^{\star}[\tau], {\hat c}_{i,n}[\tau] \right),
\end{align} 
for $i\in \{1,2,\ldots,|\mathcal{M}_{\rm t}|^{K}\}$ and $n\in \{1,2\ldots,2N\}$.

{\bf Detection:} Once the model parameter set ${\hat \Theta}[\tau]$ is updated using a new observation ${\bf y}[\tau]$, the BS performs wMHD method using the received signal for ${\bf y}[\tau]$ as
 \begin{align}
{\bf \hat x}_i[\tau]=\arg\min_{i}   d_{\sf wH}\left({\bf \hat c}_i[\tau],{\bf y}[\tau]; \{{\hat w}_{i,n}[\tau]\}_{n=1}^{2N}, \!\{{\hat w}^c_{i,n}[\tau]\}_{n=1}^{2N}\!\right). \label{eq:update_A-ML_decoding}
 \end{align}
The proposed online supervised-learning detection method is summarized in Algorithm  \ref{alg:OnlineSL_model}.

\begin{algorithm}
 	\caption{ The proposed online supervised-learning detector.}\label{alg:OnlineSL_model}
	{\small{\begin{algorithmic}[1]
					\STATE Parameter learning during the training phase: \\${\hat \Theta}[T_{\rm t}]=\left({\hat p}_{i,n}^{\star}[T_{\rm t}], {\hat c}_{i,n}[T_{\rm t}] \right)$ for $i\in \{1,2,\ldots,|\mathcal{M}_{\rm t}|^{K}\}$ and $n\in \{1,2\ldots,2N\}$.\\
					    Centroid: ${\bf \bar y}_i[T_{\rm t}] =\sum_{t\in \mathcal{T}_i}^{T_{\rm t}}{\bf y}[t]$.
	\FOR[Data detection with the updated parameters] {$t=T_{\rm t}+1,\ldots, T_{\rm B}$}	
				\STATE Compute the reliability of each codeword vector using a new observation ${\bf y}[t]$: \\$ {\hat \gamma}_i[t] = \frac{\exp\left(-d_{\sf wH}\left({\bf \hat c}_{i}[t-1],{\bf y}[t] ;\{{\hat w}_{i,n}[t-1]\}_{n=1}^{2N},\{{\hat w}^c_{i,n}[t-1]\}_{n=1}^{2N}\right)\right)}{\sum_{j=1}^{|\mathcal{M}_{\rm t}|^{K}}\exp\left(-d_{\sf wH}\left({\bf \hat c}_j[t-1],{\bf y}[t]; \{{\hat w}_{j,n}[t-1]\}_{n=1}^{2N},\{{\hat w}^c_{j,n}[t-1]\}_{n=1}^{2N}\right)\right)}$ for $i=\{1,2,\ldots, |\mathcal{M}_{\rm t}|^{K}\}$.
				\STATE Centroid update: ${\bf \bar y}_{i}[t] =  {\bf \bar y}_{i}[t-1]+{\hat \gamma}_i[t]{\bf y}[t]$ for $i=\{1,2,\ldots, |\mathcal{M}_{\rm t}|^{K}\}$.
 			    \STATE Codeword update:  ${\bf \hat c}_i[t]={\sf sign}({\bf \bar y}_i[t])$ for $i=\left\{1,2,\ldots, |\mathcal{M}_{\rm t}|^{K}\right\}$.
			    \STATE Transition probability update: $ {\hat p}_{i,n}[t] = \frac{\sum_{\tau=1}^{t}{\hat \gamma}_i[\tau]{\bf 1}_{\{ {\hat c}_{i,n}[t] \neq y_{n}[\tau]\}}}{\sum_{\tau=1}^{t} {\hat \gamma}_i[\tau]}$.
			  \STATE Detection: ${\bf \hat x}_i[t]= \arg\min_{i}   d_{\sf wH}\left({\bf \hat c}_{i}[t],{\bf y}[t] ;\{{\hat w}_{i,n}[t]\}_{n=1}^{2N},\{{\hat w}^c_{i,n}[t]\}_{n=1}^{2N}\right)$.
			\ENDFOR
	\end{algorithmic}}}
\end{algorithm}

\color{black}
 {\bf Remark 6 (Differences with the existing algorithms):} Although the proposed online parameter learning and detection method resembles the conventional EM algorithm. Unlike the conventional EM algorithm for data clustering applications, the proposed method does not iteratively perform the parameter estimation and the detection until the algorithm converges.  Instead, the proposed algorithm moves forward in a sample-by-sample fashion.  This fact facilitates to track the channel variations. In addition, the proposed online learning algorithm differs from our prior work in \cite{Jeon:WCNC:18}. The key difference is that in \cite{Jeon:WCNC:18}, the new training example set is acquired using a cyclic redundancy check (CRC) code. This method, however, requires a high computational complexity due to the iterative detection and decoding procedures. In addition, it is impossible to track the channel variation within a coding block. Whereas, the proposed method is able to instantaneously track the channel variation.  

 \section{Model-free Supervised-Learning via Deep Neural Networks} 
In this section, we present a multi-user detector using a deep neural network (DNN) by model-less supervised-learning. This approach differs from the model-based supervised-learning approaches explained in Section IV and V. For the model-less supervised-learning, we do not use any specific end-to-end network model. Specifically, the model-based approaches learn the parameters including the codebook and the set of transition probabilities that accurately matches with the likelihood function $p\left({\bf y}_j\mid {\bf x}_i; \Theta \right)$ using training examples. Using this likelihood function, the A-ML detector performs the data detection.  The deep-learning approach, however, directly learns the posteriori distribution $p({\bf x}_i\mid {\bf y}[t] ; \Theta^{\sf DNN})$ using training examples for the detection by optimizing the DNN parameters $\Theta^{\sf DNN}$. One noticeable point is that the deep-learning approach essentially does not require to know the model parameter $\Theta$ in the likelihood function.   %By using the proposed detector, we will compare the detection performance  for the data detection, 

%, we use a DNN for the data detection, in which any specific model is not exploited for parameter learning. This approach is possible to eliminate the modeling error of the parallel BSCs, in which the codebook does not capture the noise propagation effect in the multi-hop channel. 

%     \begin{figure*}[t]
% \center
%  \includegraphics[width=10cm]{DNN_figure.pdf}
%  \caption{The proposed DNN architecture.}\label{dnn_fig}
%\end{figure*}
 
%\subsection{DNN Architecture for Training and Detection}
%\begin{table}[t]
%\center
%\begin{tabular}{|c|c|c|c|c|}
%\hline
% & set1  & set2  & set3  & set4  \\ \hline
%Input Layer  & Input Symbols  & Input Symbols & Input Symbols  & Input Symbols   \\ \hline
%Hidden Layer & \begin{tabular}[c]{@{}c@{}}LSTM (50)\\ Fully Connected (30)\\ ReLU\\ Fully Connected (16)\end{tabular} & \begin{tabular}[c]{@{}c@{}}LSTM (100)\\ Fully Connected (16)\\ ReLU\\ Fully Connected (16)\end{tabular} & \begin{tabular}[c]{@{}c@{}}LSTM (50)\\ Fully Connected (16)\end{tabular} & \begin{tabular}[c]{@{}c@{}}LSTM (100)\\ Fully Connected (16)\end{tabular} \\ \hline
%Output Layer & Softmax                                                                                                & Softmax                                                                                                 & Softmax                                                                  & Softmax                                                                   \\ \hline
%\end{tabular}
%\caption{Various deep neural network structures.} \label{dnn_set}
%\end{table}

\begin{table}[t]\label{dnn_set}
\center
 
\begin{tabular}{c|c|c|c|c}
\hline
 Layer & Type & Size  & Activation  & Weight parameters  \\ \hline\hline
Input Layer  & Received signals (Labeled)  & $2N$ & $-$  & $-$   \\ \hline
Hidden Layer-1  & LSTM  &  $4((2N+1)\rho +\rho^2)$ & ReLU  & $\Theta_{1}^{\sf DNN}$   \\ \hline
Hidden Layer-2  & Fully connected  &  $|\mathcal{M}_{\rm t}|^{K}(\rho + 1)$ & $-$  & $\Theta_{2}^{\sf DNN}$   \\ \hline
Output Layer  & Fully connected  &  $|\mathcal{M}_{\rm t}|^{K}$ & Softmax  & $\Theta_{3}^{\sf DNN}$   \\ \hline
\end{tabular}\vspace{-0.3cm}
\caption{ The used DNN structures. Here, $\rho$ is the number of output parameters for LSTM and the DNN parameter set is denoted by $\Theta^{\sf DNN}=\left\{\Theta_{1}^{\sf DNN},\Theta_{2}^{\sf DNN},\Theta_{3}^{\sf DNN}\right\}$. } 
\end{table}

{\bf The proposed DNN architecture:} We construct a DNN with multiple layers to capture the multi-hop channels, one-bit quantization, and noise effects. As illustrated in Table I, the proposed DNN architecture is composed of an input layer, two hidden layers, and an output layer. The input layer takes $2N$ dimensional binary received signals ${\bf y}[t]\in \{-1,1\}^{2N}$. Since this input signal of the DNN is spatially correlated due to the multi-hop MIMO channels, a long short-term memory (LSTM) network is used for the first hidden layer to exploit this correlation structure in the received signal. We also employ a fully connected network for the second hidden layer. Then the output layer computes the posteriori distribution using the softmax function. Let $u_{i}[t](\Theta^{\sf DNN})$ be the $i$th output of the output layer at the $t$th training sample, which is a function of the DNN parameters $\Theta^{\sf DNN}$. Then, the $k$th output of the softmax function is given by
\begin{align}
    p_k[t] (\Theta^{\sf DNN})= \frac{e^{u_i[t](\Theta^{\sf DNN})}}{\sum_{i=1}^{|\mathcal{M}_{\rm t}|^{K}}e^{u_i[t](\Theta^{\sf DNN})}}.
  %  \mathbb{P}\left[{\bf x}[t]={\bf x}_k \mid  {\bf y}[t]~;\Theta^{\sf DNN} \right].
\end{align}

%\begin{align}
%	u_i[t] = \frac{e^{o_k} }{\sum_{j=1}^{2^{2K}}e^{o_i} }
%\end{align}

As can be seen, the output of the softmax function is the probability that the received signal belongs to the $k$th information vector. 

%This estimated probability mass function is used for a soft decision in detecting the symbol. The overall DNN architecture is illustrated in Fig. \ref{dnn_fig}.
%\begin{align}
%\mathbf{f}_t &= \sigma \left(\mathbf{W}_f\mathbf{y}_t + \mathbf{U}_f\mathbf{h}_{t-1}+\mathbf{b}_f \right) \nonumber \\
%    \mathbf{i}_t &=\sigma \left(\mathbf{W}_i \mathbf{y}_t + \mathbf{U}_i \mathbf{h}_{t-1} +\mathbf{b}_i\right) \nonumber \\
%    \tilde{\mathbf{c}}_t &= \text{tanh} \left( \mathbf{W}_c \mathbf{y}_t+\mathbf{U}_c \mathbf{h}_{t-1}+\mathbf{b}_c\right) \nonumber \\
%    \mathbf{c}_t &= \mathbf{f}_t \otimes \mathbf{c}_{t-1} + \mathbf{i}_t \otimes \tilde{\mathbf{c}}_t \nonumber \\
%    \mathbf{o}_t &= \sigma \left(\mathbf{W}_o \mathbf{y}_t + \mathbf{U}_o \mathbf{h}_{t-1}+\mathbf{b}_o \right) \nonumber \\
%    \mathbf{h}_t &= \mathbf{o}_t \otimes \text{tanh}(\mathbf{c}_t),
%\end{align}
%where $\sigma (\cdot)$ is the logistic sigmoid function and $\otimes$ is the Hadamard product. The output of the LSTM layer is $\mathbf{h}_t$. 

{\bf Training the DNN detector:} During the training phase, the DNN detector is trained to classify the received signal as one of the possible transmit symbols in $\mathcal{X}$. Specifically, by sending pilot symbols, the BS obtains $T_{\rm t}=T|\mathcal{M}_{\rm t}|^{K}$ labeled training examples $\mathcal{S}=\left\{ ({\bf x}[1],{\bf y}[1]), \ldots, ({\bf x}[T_{\rm t}],{\bf y}[T_{\rm t}])\right\}$. The DNN is trained to minimize a loss function between the outputs of the DNN and the labeled data.  The cross-entropy loss function is used  for the parameter optimization as
\begin{align}
    \mathcal{L}( \Theta^{\sf DNN}) = -\sum_{t=1}^{T_{\rm t}}\sum_{i=1}^{|\mathcal{M}_{\rm t}|^{K}}\mathbf{1}_{\left\{\mathbf{x}[t]=\mathbf{x}_i\right\}}\ln  p_k[t] (\Theta^{\sf DNN}).
\end{align}
The well-known stochastic gradient method is used to estimate the parameters of the DNN.
%Unlike the model-based approaches explained in the previous sections, in which the parameters of the likelihood function are learned, i.e., $\mathbb{P}\left[ {\bf y}[t]={\bf y}_j | {\bf x}[t]={\bf x}_i; \Theta\right]$, the DNN learns the posterior probability distribution by optimizing the loss function, namely, 
%\begin{align}
%	\mathbb{P}[{\bf x}[t]={\bf x}_i \mid {\bf y}[t]; \{\Theta_{k}^{\sf DNN}\}_{k=1}^3].
%\end{align}  

{\bf Detection:} Detecting the transmit symbol from the received signal can be a classification problem when the number of possible transmit symbols is finite. In our problem, there are $|\mathcal{M}_{\rm t}|^{K}$ classes from $1$ to $|\mathcal{M}_{\rm t}|^{K}$ that are corresponding to the $|\mathcal{M}_{\rm t}|^{K}$ possible transmit symbols. Once the network parameters are learned using the training examples, the DNN detector performs the detection for the received signal vector during the data detection phase, i.e., ${\bf y}[t]$ for $t\in\{T_{\rm t}+1,\ldots, T_{\rm B}\}$.

%Then the received vector $\mathbf{y}[t]$ is classified as one of the $|\mathcal{X}|=M^{K}$ classes and that class is mapped into the corresponding transmit signal. 
%
%By using the stochastic gradient algorithm, the parameters of the DNN are optimized so as to minimize the cross-entropy loss function, i.e.,
%\begin{align}
%    \mathcal{L} = -\sum_{t=1}^{T_{\rm t}}\sum_{i=1}^{|\mathcal{X}|}\mathbf{1}_{\left\{\mathbf{x}[t]=\mathbf{x}_i\right\}}\ln p_i[t],
%\end{align}
%where $p_i[t]$ is the output value from the softmax function. 

{\bf Remark 7 (Practical challenges for the DNN detector):} This model-free approach is useful in improving the detection performance by eliminating the possible model errors. Nevertheless, the DNN detector is not suitable for the scenarios where the channel changes relatively fast, because the computational complexity for learning the weight parameters is very high compared to the model-based supervised-learning approach.

\color{black}
\section{Simulation Results}
 
In this section, we evaluate symbol-error-rate (SER) and the symbol-vector-error-probability (SVEP) performances for classical, model-based, and model-free approaches. In our simulation, we consider Rayleigh-fading channels, in which each element of the channel matrix per hop is drawn from IID complex Gaussian random variables, i.e., $\mathcal{NC}\left(0,1\right)$. In addition, we consider the two-hop MU-MIMO system in which the SNR of the first-hop is fixed, while the second-hop SNRs are changed. For a notation, we simply denote the two-hop MU-MIMO channel with $K$ uplink users, $L_1$ distributed relays, and $N$ BS antennas by $[K,L_1,N]$.

%For the detectors of the classical approach, we consider two linear detectors

% 
%  \begin{itemize}
%      \item ZF : For the ZF detector, we estimate the data symbol as
%          \begin{align}
%          \hat{\mathbf{x}}^{\text{ZF}} = (\mathbf{H}_1^{\sf H}\mathbf{H}_1)^{-1} \mathbf{H}_1^{\sf H} (\mathbf{H}_2^{\sf H}\mathbf{H}_2)^{-1} \mathbf{H}_2^{\sf H} \mathbf{y}.
%          \end{align}
%      \item LMMSE : For the LMMSE detector, we estimate the data symbol as
%           \begin{align}
%          &\hat{\mathbf{x}}^{\text{LMMSE}} \nonumber \\&= \mathbf{H}_1^{\sf H}(\mathbf{H}_1\mathbf{H}_1^{\sf H} + \mathbf{\sigma}_{v_1}^{2}\mathbf{I} )^{-1} \mathbf{C_{r'}}\mathbf{H}_2^{\sf H} (\mathbf{H}_2 \mathbf{C_{r'}}\mathbf{H}_2^{\sf H} + \mathbf{\sigma}_{v_2}^{2}\mathbf{I})^{-1} \mathbf{y},
%           \end{align}
%           where $\mathbf{C_{r'}}$ denotes the auto-correlation matrix of the vector $\mathbf{H}_1\mathbf{x}+\mathbf{v}_1$.
%  \end{itemize}

\begin{figure}[t]
    \centering
\epsfig{file=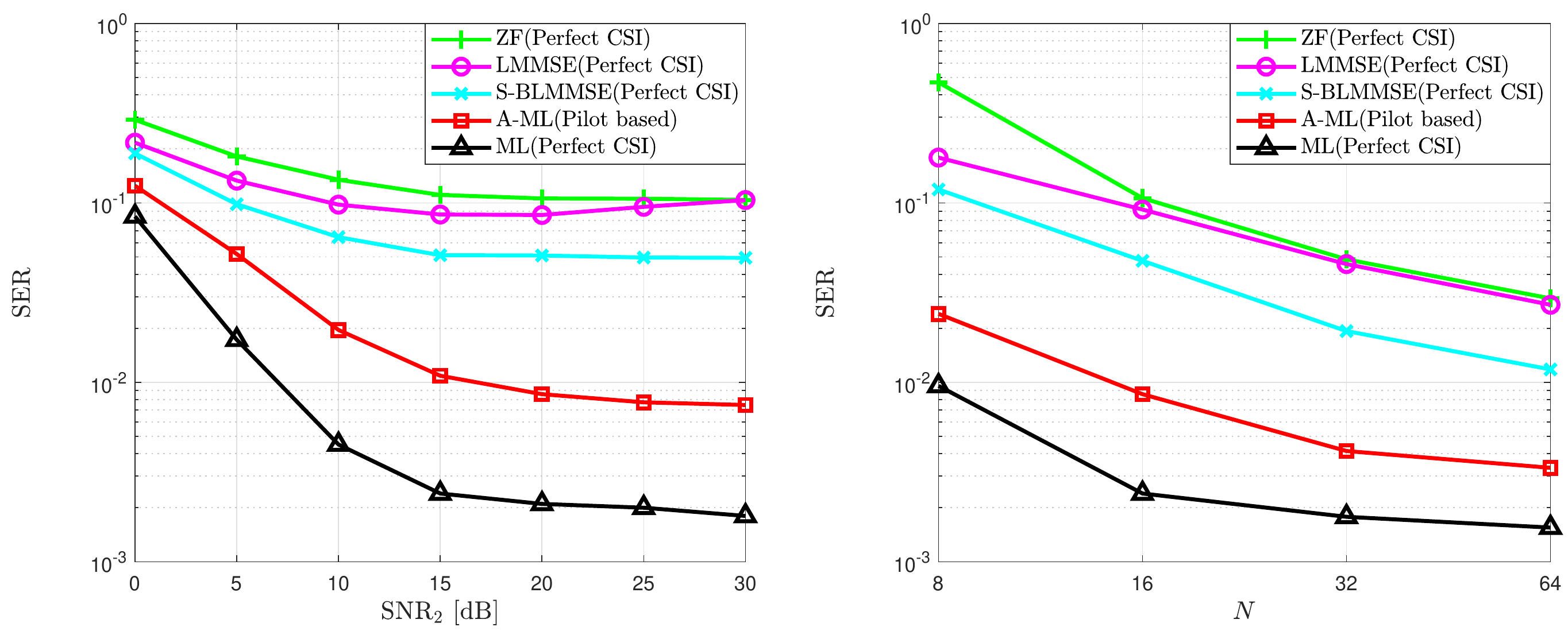,width=16.0cm}\vspace{-0.3cm}
\caption{ The SER comparison for $[K,L_1,N]=[2,8,16]$ (left) and $[K,L_1,N]=[2,8,N]$ (right). The QPSK modulation is used per user.}
\label{scheme_plot}
\end{figure}

Fig. \ref{scheme_plot} compares the SER performances between conventional detectors based on the classical approach and the A-ML detector based on the model-based supervised learning approach. In addition to the ML detector, we consider three linear detectors including a zero-forcing detector, a linear minimum mean square error (LMMSE) detector, and a successive Bussgang linear MMSE (S-BLMMSE) detector. In particular, the S-BLMMSE detector is designed by successively applying the Bussgang decomposition in \cite{Li}. Unlike the conventional detectors in which perfect and global CSI is available at the BS, the proposed A-ML detector uses the estimated model parameters by sending $T=15$ pilots per information symbol vector. The first-hop SNR is fixed to 20dB. As can be seen in the left figure, it is observed that the proposed A-ML detector significantly outperforms the linear detectors, even with imperfect knowledge of the model parameters. The similar performance tendency is observed when increasing the number of antennas at the BS, when the second-hop SNR is fixed to 20dB.

%Fig. \ref{scheme_plot} demonstrates the SER performances the first-hop SNR is set to be 20dB.  We 
% The left figure shows the SER performances where $[K,L_1,N] =[2,8,16]$ and the right figure shows the SER performances for the various number of BS antennas with 20dB of second-hop SNR. 

%We set the training length of $T=15$ for the A-ML detector.  For the zero forcing (ZF) detector and linear minimum mean square error (LMMSE) detector, we apply the two-stage ZF and LMMSE detection method. Further, we use a successive Bussgang linear MMSE (S-BLMMSE) where the Bussgang decomposition is applied successively to linearize the quantization systems. As can be seen, the proposed ML detector achieves the optimal SER performance, which requires the perfect and global knowledge of CSI at the BS. Whereas, using a limited number of channel training overheads the proposed A-ML detector significantly reduces the computational complexity compared to that of the ML. In addition, the proposed A-ML method considerably outperforms the linear detection methods including ZF, LMMSE and the S-BLMMSE, which require the global CSI at the BS. 

\begin{figure}[t]
    \centering
\epsfig{file=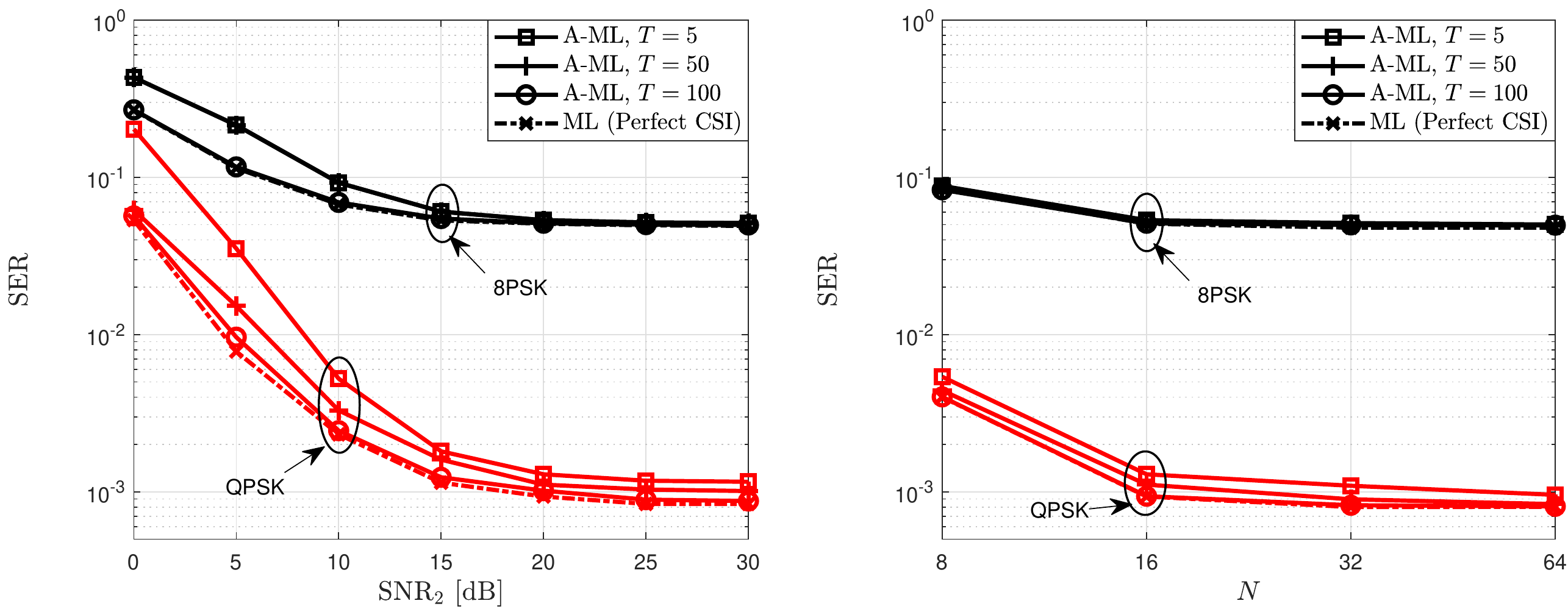,width=16.0cm}\vspace{-0.3cm}
\caption{ The SER comparison of the A-ML detector for $[K,L_1,N]=[2,8,16]$ (left) and $[K,L_1,N]=[2,8,N]$ (right). The QPSK and 8PSK modulations are considered.}\label{training_length}
\end{figure}

% Fig. \ref{scheme_plot} demonstrates the SER performances for the two-hop multi-user MIMO system where [$K,L_1,N$]=[2,8,64]. The first-hop SNR is fixed to 20dB. We set the training length of $T=15$ for the A-ML detector. As can be seen, the proposed ML detector achieves the optimal SER performance, which requires the perfect and global knowledge of CSI at the BS. Whereas, using a limited number of channel training overheads the proposed A-ML detector significantly reduces the computational complexity compared to that of the ML. In addition, the proposed A-ML method considerably outperforms the linear detection methods including ZF, LMMSE and the proposed S-BLMMSE, which require the global CSI at the BS. The propose S-BLMMSE is better in terms of the SER compared to the conventional linear detection methods including ZF and LMMSE. 

% {\bf  Optimality of the proposed A-ML detection:} To verify the claim in Theorem 1, we compare the SER performance

Fig. \ref{training_length} shows the SER performances of the A-ML detector as $T$ increases. In particular, to verify Theorem 1, we compare the SER performances between the ML and the A-ML detectors by increasing $T$, assuming the first-hop SNR is infinite, $\sigma_{1}=0$.  As can be seen in Fig. \ref{training_length}, the SER performance gap between the proposed A-ML and the ML detectors diminishes by increasing $T$. This result agrees with Theorem 1. One remarkable observation is that the A-ML detector achieves a near-ML performance even with a reasonable amount of pilots when the second-hop SNR is beyond 15 dB.

\begin{figure} [t]
	\centering
    \includegraphics[width=9cm]{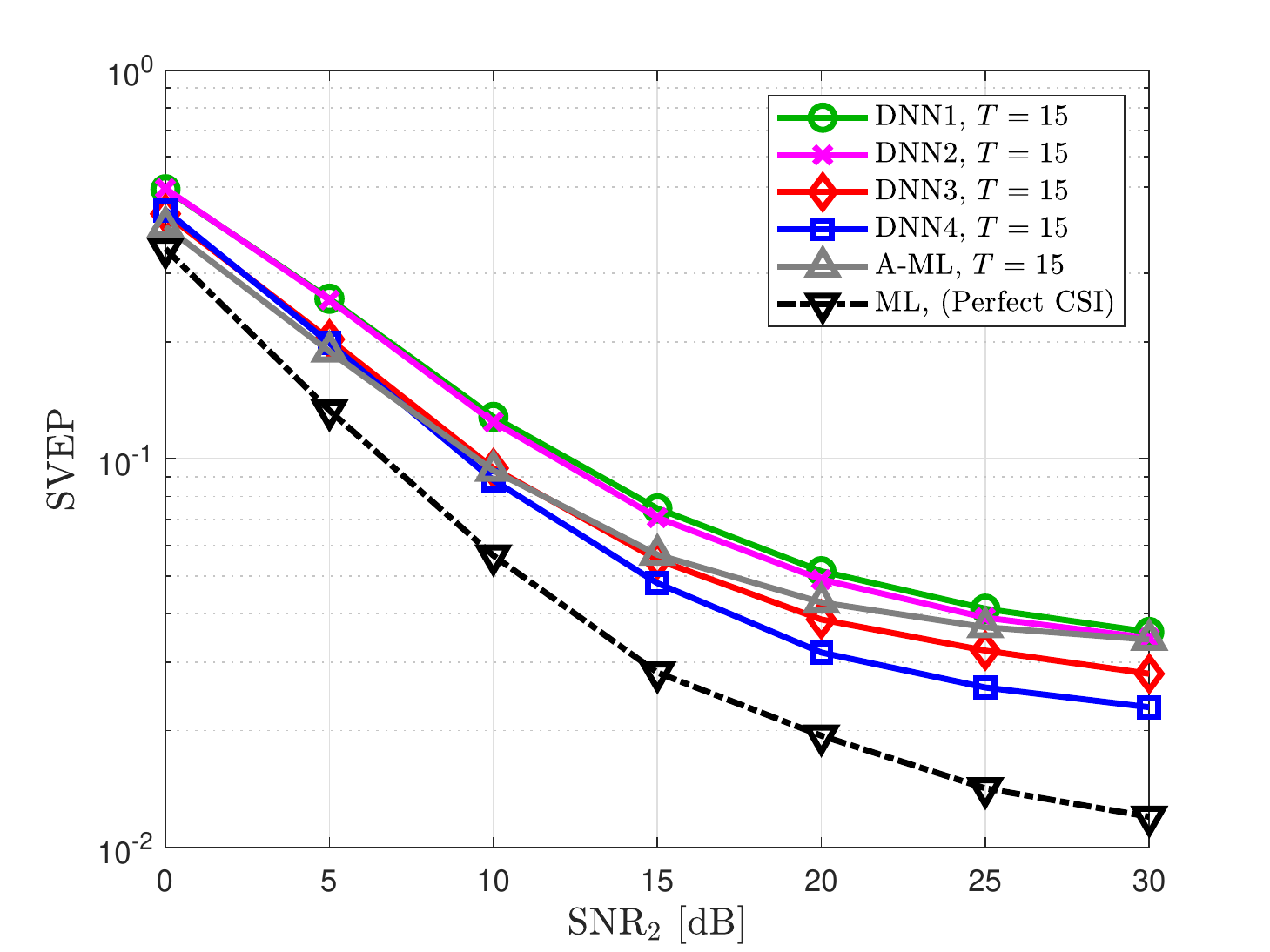}
  \caption{The SVEP comparison for the different DNN configurations when QPSK modulation is used where $[K,L_1,N]=[2,8,8]$ and SNR$_1$=20dB.} \label{DNN} 
\end{figure}

We evaluate the SVEP performance for several DNN detectors using various configurations and parameter settings as shown in Table \ref{dnn_set}. As shown in Fig. \ref{DNN}, when the number of the layers increases, the performance degrades because of the overfitting problem. On the contrary, when the number of layers is fixed, we observe that the use of a sufficient number of hyper-parameters per layer performs better. Therefore, we use the DNN4 detector to compare with the other schemes including the A-ML and ML detectors.   
\begin{table}[h]
\centering
\resizebox{\textwidth}{!}{
\begin{tabular}{c|c|c|c|c}
\hline
& DNN1 & DNN2 & DNN3 & DNN4 \\ \hline
Input Layer & Received signals & Received signals & Received signals & Received signals \\\hline
Hidden Layer & \begin{tabular}[c]{@{}c@{}}LSTM (50)\\ Fully Connected (30)\\ ReLU\\ Fully Connected (16)\end{tabular} & \begin{tabular}[c]{@{}c@{}}LSTM (100)\\ Fully Connected (16)\\ ReLU\\ Fully Connected (16)\end{tabular} & \begin{tabular}[c]{@{}c@{}}LSTM (50)\\ReLU\\ Fully Connected (16)\end{tabular} & \begin{tabular}[c]{@{}c@{}}LSTM (100)\\ ReLU\\Fully Connected (16)\end{tabular} \\ \hline
Output Layer & Softmax & Softmax & Softmax & Softmax \\ \hline
\end{tabular}}
\vspace{0.2cm}
\caption{Different configurations of the DNN detectors.} \label{dnn_set}
\end{table}

\begin{figure}[t]
    \centering
\epsfig{file=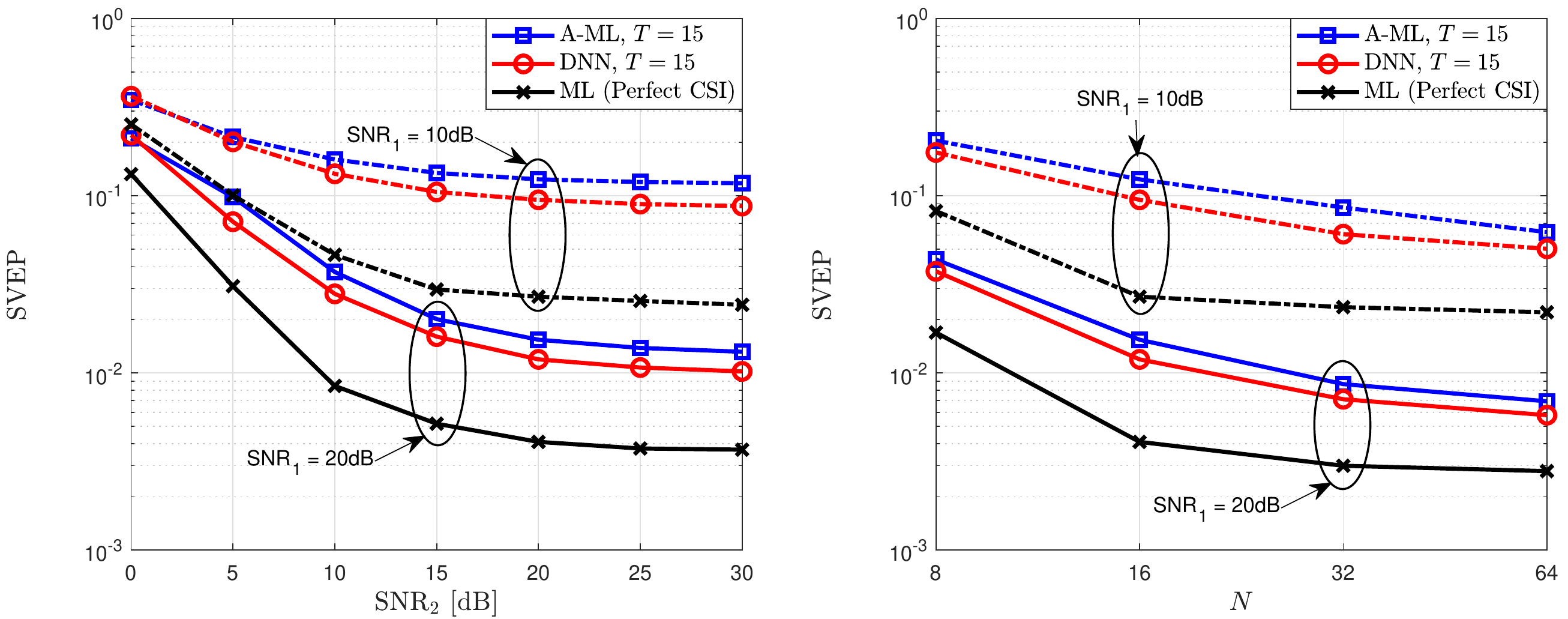,width=16.5cm}\vspace{-0.3cm}
\caption{ The SVEP comparison for the proposed detectors for $[K,L_1,N]=[2,8,16]$ (left) and $[K,L_1,N]=[2,8,N]$ (right). The QPSK modulation is used.}\label{dnn_plot}
\end{figure}

Fig. \ref{dnn_plot} compares the SVEP performances for three different detection approaches: the classical, the model-based, and the model-free approaches. In this simulation, we evaluate the SVEP performances at two different SNRs of the first-hop. In addition, we train the parameters for the DNN and the A-ML detectors per SNR by sending the pilots with the length of $T=15$. In particular, for the DNN detector, we chose $\rho=100$ for the LSTM layer \footnote{ We simulate using different values of $\rho$ to train the DNN detector. We observe that the detection performances are similar when $100\leq \rho \leq 150$. However, the performance is degraded when $\rho \geq 150$ due to a overfitting problem.}. One interesting observation is that the proposed DNN detector slightly outperforms the A-ML detector, because it is capable of removing the model errors. This also happens when increasing the number of antennas at the BS, when the second-hop SNR is fixed to 20dB. Nevertheless, the computational complexity of the DNN detector is much higher compared to that of the A-ML detector. For instance, for a given channel realization and one SNR point, the runtimes for the DNN,  the ML and the A-ML detection algorithms are measured as 11.383sec, 2.052sec, and 0.0931sec, respectively under the same simulation condition. This result is because the DNN detector uses a more number of hyper-parameters than that of the A-ML detector as shown in Table I.

 \begin{figure}[t]
    \centering
\epsfig{file=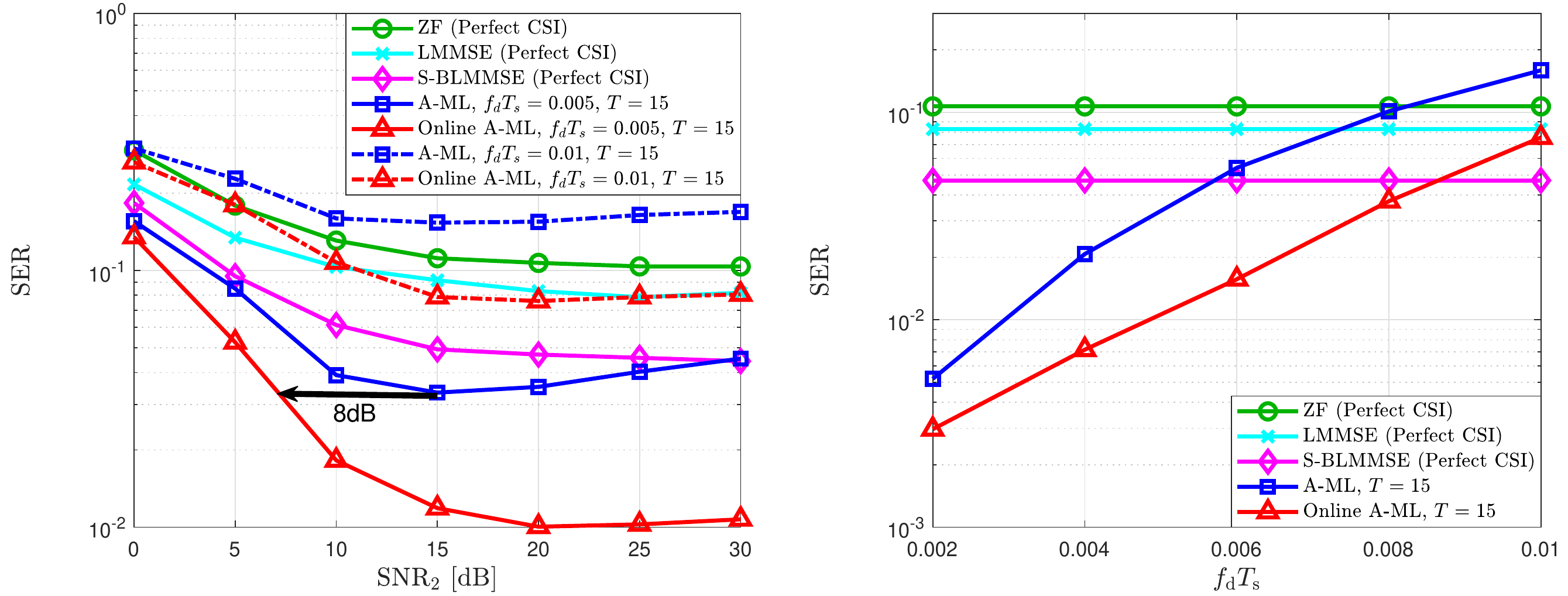,width=16.5cm}\vspace{-0.3cm}
\caption{ The SER comparison for the detectors under the time-varying channels. The channel configuration is $[K,L_1,N]=[2,8,16]$ and the QPSK modulation is used.}\label{online}
\end{figure}
Fig. \ref{online} compares the SER performances of various detectors in time-varying channels. In this simulation, we assume that the second-hop channel is time-varying, while the first-hop channel is  time-invariant.  To model the time-varying channel of the second-hop, we use an order-one auto-regressive process as $
	{\bf H}_2[t]=\eta{\bf H}_2[t-1]+{\bf W}[t],$ where $\eta$ is the temporal correlation coefficient for the second-hop channel fading and ${\bf W}[t]$ is a process noise matrix whose $(i,j)$ element is drawn from a complex Gaussian random variable, i.e., ${\bf W}_{i,j}[t]\sim \mathcal{CN}(0,1-\eta^2)$. Using the Jakes's model, the temporal correlation coefficient is chosen as  $\eta=J_0(2\pi f_{\rm d}T_{\rm s})$, where $J_0(\cdot)$ denotes the Bessel function of the first kind of order zero, $f_{\rm d}$ is the maximum Doppler frequency, and $T_{\rm s}$ is the sampling time. In our simulation, we assume that the channel is invariant during the training phase, and the SNR of the first hop is 30dB. As can be seen in Fig. \ref{online}, when $f_{\rm d}T_{\rm s}=0.005$, the online-learning based A-ML detector significantly outperforms the existing linear detectors, which use the global and perfect CSI knowledge at the BS. In addition, it is shown that the proposed online supervised-learning detector provides a considerable SER gain over the A-ML detector that does not update the model parameters in a symbol-by-symbol fashion. For example, the online supervised-learning detector yields about 8dB SNR gain over the supervised-learning detector for a target SER of 0.03. When increasing the normalized doppler $f_{\rm d}T_{\rm s}$, however, the SER performance is degraded, because the larger $f_{\rm d}T_{\rm s}$ causes the more channel variation; thereby, it makes harder to track the channel variation. Nevertheless, the performance of the online supervised-learning detector is similar to that of the LMMSE detector that uses the global and perfect CSI knowledge at the BS.  %This result confirms that the proposed online learning based detector can be useful in practical systems.

\color{black}\section{Conclusion}

In this paper, we introduced a new nonlinear MU-MIMO relay channel, in which distributed relays use one-bit DACs and ADCs motivated by low-power hardware constraints. In this channel, to understand the limit of the multi-user detection performance, we first proposed the ML detector, which requires global and perfect CSI at the BS. Inspired by an end-to-end supervised-learning technique, we presented a novel data communication framework by developing a simple yet effective channel model. The proposed model facilitated learning parameters using a simple pilot transmission strategy, while ensuring the optimality of the detection performance in some conditions. In addition, we extended the proposed communication framework into a time-varying channel environment. The proposed online supervised-learning detector jointly performed the update of the model parameters and the data detection using unlabeled received signals via the EM-like algorithm. Lastly, we also presented a detector using a DNN that does not rely on any specific network model.  Via simulations, we compared the SER performances of different detection approaches in order to provide a complete view on the effectiveness of using supervised-learning in the considered MIMO channel. 

%It would be interesting to extend the proposed supervised-learning approaches for frequency-selective channel models and channel coding techniques in the nonlinear multihop relay channels using results in \cite{SN_Hong:18,Jeon:TC:18}. 

\end{document}